\documentclass[aps,pra,twocolumn,floatfix]{revtex4-2}
\usepackage{amsmath,amssymb,graphicx,bbold}
\usepackage{graphicx}  
\usepackage{dcolumn}   
\usepackage{bm}        
\usepackage{color}
\usepackage{xcolor}

\usepackage{xcolor}
\usepackage{physics}
\usepackage{ulem}
\usepackage{subfigure}

\begin{document}

\title{Algebraic approach to a nonadiabatic coupled  Otto cycle}

\author{A. C. Duriez, D. Mart\'{i}nez-Tibaduiza, A. Z. Khoury}
\affiliation{Instituto de F\'{i}sica, Universidade Federal Fluminense, 24210-346 Niter\'{o}i, RJ, Brazil}
\date{\today}

\begin{abstract}
Algebraic methods for solving time-dependent Hamiltonians are used to investigate the performance of quantum thermal machines.
We investigate the thermodynamic properties of an engine formed by two coupled qubits, performing an Otto cycle. The thermal interaction occurs with two baths at different temperatures, while work is associated with the interaction with an arbitrary time-dependent magnetic field that varies in intensity and direction. For the coupling, we consider the 1D isotropic Heisenberg model, which allows us to describe the system by means of the irreducible representation of the $\mathfrak{su}(2)$ Lie algebra within the triplet subspace. We also inspect different settings of the temperatures and frequencies of the cycle, and investigate the corresponding operation regimes. We describe how these regimes can change with the speed of the protocol.
Finally, we compare the weak and strong coupling regimes, and conclude that for the weak case the efficiency is optimal and can surpass the value of the classical Otto cycle. 
For low temperatures of the heat baths, we show that losses due to quantum friction are reduced in the strong coupling limit.
\end{abstract}

\maketitle
\section{Introduction}
\label{intro}

The theory of classical thermodynamics is, without question, one of the most successful in all of physics. One of the beauties of the subject is that it can describe an immense range of phenomena using only a reduced number of parameters, like the entropy, temperature and volume \cite{Callen}.
The success of the phenomenological descriptions provided by this formalism was so great that, in the present day, the laws of thermodynamics occupy a higher place in the hierarchy of physical theories, in a sense that all new theories must agree with the established principles of conservation of energy and entropy increase, for example.

The quest for new quantum devices requires a suitable description of their microscopic components, especially 
the operations performed on them.
The field of quantum thermodynamics has emerged in the recent time, partially, as an attempt to recover, or reformulate, the laws of thermodynamics and some concepts like, heat, work and entropy production by starting within a full quantum perspective. Not surprisingly, the analysis of driven quantum systems in contact with thermal environments has led to interesting results when made through a thermodynamic inspired framework. In 1959, Scovil and Schulz-Dubios described how a three level maser can be equivalent to a Carnot Engine, in a work that today is considered a seminal paper \cite{3levelmaser}. 
Motivated by this and other important results, the study of quantum thermal machines has flourished. Today, there is a great deal of theoretical and experimental work devoted to improve the performance of microscopic machines.

A spin system controlled by an external magnetic field constitutes a possible realization of a quantum thermal machine. Considering the spins as the working substance, protocols where the external magnetic field changes in time can be considered as work protocols in the quantum thermodynamics framework. We mention a few of the many references where quantum thermal machines of one and two spins are analyzed \cite{CBPF1,CBPF2,Kosloff-FourStroke,article}.

In this work we investigate the thermodynamic properties of a two spin coupled system performing an Otto cycle, and we use an algebraic approach to describe the effects of finite-time transitions in the unitary strokes.
In Sec. \ref{sec:algebra} we present the algebraic method that is used in the numerical calculations of the unitary evolutions \cite{NewBCH}. In Sec. \ref{section:model} we use the algebraic methods to expand the formalism presented in \cite{Campisi1} to the case of two interacting spins, where we describe the interaction by the isotropic Heisenberg model and consider the action of an external magnetic field with a time varying amplitude and direction. In \cite{2qbitOttoengine1}, the authors consider the same system, but they restrict their analysis to the adiabatic case. In Sec. \ref{section:twoqbitottocycle}, we describe the Otto cycle for the working substance considered, and we compute the heat exchanged with the hot and cold sources as well as the output work per cycle in terms of the persistence probabilities of the instantaneous eigenstates of the Hamiltonian. In Sec. \ref{sec:Operation}, we describe the dynamics of a specific protocol of the driving field  and we investigate regions of parameters where the machine operates in the four possible regimes of heater, refrigerator, accelerator and engine. We also describe how the machine can transition between different operation regimes as the rate of change of the driving field, associated with the degree of adiabaticity of the protocol, increases. In Sec. \ref{sec:Performance_Optimization}, we focus on the engine operation mode, and we analyze the performance in the limits of weak and strong coupling. We show that the efficiency can surpass the bound of the classical Otto cycle  in the weak coupling regime. For a strong coupling, we show that the energy losses due to friction can be suppressed for low temperatures of the heat baths. Sec. \ref{section:conclusion} is left for conclusions and prospects for future analyses. 

\section{Algebraic solution for time-dependent Hamiltonians}
\label{sec:algebra}

In this section we discuss the algebraic method presented in \cite{NewBCH} to describe the dynamics of quantum 
systems with Hamiltonians that are linear combinations of generators of the $\mathfrak{su}(2)$, $\mathfrak{su}(1,1)$ or $\mathfrak{so}(2,1)$ Lie algebras. This method will be useful in our future analysis of quantum thermal machines, where they will be applied to the unitary strokes of the quantum thermal cycle. 
In quantum mechanics, unitary operators are often written as the exponential of a linear combination of operators that can or not commute. Moreover, for the noncommuting case, the factorization is quite non trivial, as stated in \cite{Barnett-Book-1997}. However, when the operators in the exponent are generators of some particular Lie algebra, there is an elegant way to proceed. Consider the operators satisfying 
\begin{equation}
\left[T_{-},T_{+}\right]=2\epsilon T_{c}, \:\:\: \:\:\: \left[ T_{c}, T_{\pm}\right]=\pm\delta T_{\pm}\, .
\label{eq:algebraK}
\end{equation}
The commutation relations for the $\mathfrak{su}(2,1)$, $\mathfrak{su}(2)$ and $\mathfrak{so}(2,1)$ Lie algebras fall in this general form. Parameters $\epsilon$ and $\delta$ are introduced in order to treat the three algebras in the same framework, and their values for the given algebras are indicated in Table \ref{tab:I}.
\begin{table}[h!]
\centering
\begin{tabular}{ | m{6em} | m{0.8cm}| m{0.4cm} | } 
 \hline
Lie Algebra & $\delta$ & $\epsilon$ \\
 \hline
\textit{su}$(1,1)$  & $1$  & $1$  \\
\hline
 \textit{su}$(2)$  & $1$  & $-1$ \\
\hline
\textit{so}$(2,1)$  & $i$  & $i/2$ \\
\hline
\end{tabular}
		\caption{Relation between the Lie algebras under consideration and parameters $\delta$ and $\epsilon$.}
\label{tab:I}
\end{table}
Now consider the operator
\begin{align}
    G=f(\boldsymbol{\lambda})=e^{\lambda_+ T_++\lambda_c T_c+\lambda_- T_-}, \label{eq:Goperator1}
\end{align}
where $\boldsymbol{\lambda}=(\lambda_+,\lambda_c,\lambda_-)$ is a set of complex parameters. Since the $ T$ operators are  generators of the given algebras, the $G$ operators are elements of the group associated with those algebras, by definition of the group structure \cite{gilmore2012lie}.
For the algebras presented here, it can be shown that any group element as in Eq. \eqref{eq:Goperator1} can also be written in a factorized form,
\begin{align}
    G=h(\boldsymbol{\Lambda})=e^{\Lambda_+ T_+}e^{\ln{\Lambda_c} T_c}e^{\Lambda_- T_-}. \label{eq:factorization}
\end{align}
The new set of parameters $\boldsymbol{\Lambda}$ is related to the old one by
\begin{eqnarray}
\Lambda_{c} &=& \left(\cosh(\nu)-\frac{\delta\lambda_{c}}{2\nu}\sinh(\nu)\right)^{-\frac{2}{\delta}}\;,
\nonumber\\
\Lambda_{\pm} &=& \frac{2\lambda_{\pm}\sinh(\nu)}{2\nu \cosh(\nu)-\delta\lambda_{c}\sinh(\nu)}\, , 
\label{eq:biglambdas}
\end{eqnarray}
with $\nu$ given by 
\begin{equation}
\label{eq:nu}
\nu^{2} = (\frac{\delta\lambda_c}{2})^2-\delta\epsilon\lambda_{+}\lambda_{-} \,\, .
\end{equation}
A proof of this result for the cases of the $\mathfrak{su}(2)$ and the  $\mathfrak{su}(1,1)$ algebras  
is given in \cite{twojumps}, and the generalization for the other case is presented in \cite{NewBCH}. The relations between these sets of parameters are known in the literature as the BCH-like relations, in reference to the Baker-Campbell-Hausdorff relations, which are used in the derivation of these results.


\subsection{The New BCH-like relations}
\label{subsection:TNewBCH}

Another result that will be useful to our future analysis is the composition of $N$ group elements of the given algebras \cite{NewBCH}. 
More specifically, we want to write down the rule that allows one to find the resulting element of the 
arbitrary composition of $G$ operators given in the factorized representation of Eq. \eqref{eq:factorization}. 
To do this, we need first to calculate how two elements compose. Such composition reads
\begin{eqnarray}
&&G(\boldsymbol{\Lambda}_2) G(\boldsymbol{\Lambda}_{1})=
\label{eq:compo}\\
&&e^{\Lambda_{2+} T_{+}}e^{\ln(\Lambda_{2c}) T_{c}}e^{\Lambda_{2-} T_{-}}e^{\Lambda_{1+} T_{+}}e^{\ln(\Lambda_{1c}) T_{c}}e^{\Lambda_{1-} T_{-}}\,.
\nonumber
\end{eqnarray}
This expression can be cast as a single $G$ operator when, using ordering techniques, 
we group together operators with the same generator in the exponent, namely, 
\begin{equation}
 G (\boldsymbol{\Lambda}_{2}) G (\boldsymbol{\Lambda}_{1})=e^{\alpha_2 T_+}e^{\ln{(\beta_2)} T_c}e^{\gamma_2 T_-}, \label{eq:newBCH12}
\end{equation}
with
\begin{align}
    \alpha_{2}&=\Lambda_{2+}+\frac{\Lambda_{1+}(\Lambda_{2c})^\delta}{1-\epsilon\delta\Lambda_{1+}\Lambda_{2-}} , \nonumber \\ \beta_2&=\frac{\Lambda_{1c}\Lambda_{2c}}{(1-\epsilon\delta\Lambda_{1+}\Lambda_{2-})^\frac{2}{\delta}}, \label{eq:newBCH1} \\
 \gamma_2&=\Lambda_{1-}+\frac{\Lambda_{2-}(\Lambda_{1c})^\delta}{1-\epsilon\delta\Lambda_{1+}\Lambda_{2-}}. \nonumber
\end{align}
In the same way, composing a third element of the group from the left of Eq. \eqref{eq:newBCH12} yields
\begin{equation}
 G (\boldsymbol{\Lambda}_{3}) G (\boldsymbol{\Lambda}_{2}) G (\boldsymbol{\Lambda}_{1})=G(\Lambda_{3+},\Lambda_{3c},\Lambda_{3-})G(\alpha_{2},\beta_{2},\gamma_{2}). \label{eq:newBCH2}
\end{equation}
As we can see, the right-hand side of the equation above is just another composition of two $ G $ operators. We can therefore use Eqs. \eqref{eq:newBCH12} and \eqref{eq:newBCH1} to write
\begin{equation}
 G (\boldsymbol{\Lambda}_{3}) G (\boldsymbol{\Lambda}_{2}) G (\boldsymbol{\Lambda}_{1})=e^{\alpha_3 T_+}e^{\ln{(\beta_3)} T_c}e^{\gamma_3 T_-}, \label{eq:newBCH3}
\end{equation}
with
\begin{align}
    \alpha_{3}&=\Lambda_{3+}+\frac{\alpha_{2}(\Lambda_{3c})^\delta}{1-\epsilon\delta\alpha_{2}\Lambda_{3-}}\,, 
    \nonumber\\
    \beta_3&=\frac{\beta_{2}\Lambda_{3c}}{(1-\epsilon\delta\alpha_{2}\Lambda_{3-})^\frac{2}{\delta}}\,,  
    \label{eq:newBCH4}\\
    \gamma_3 &=\gamma_{2}+\frac{\Lambda_{3-}(\beta_{2})^\delta}{1-\epsilon\delta\alpha_{2}\Lambda_{3-}}\,.
    \nonumber
\end{align}
By inspection of Eqs. \eqref{eq:newBCH1} and \eqref{eq:newBCH4}, we can see that the coefficients $(\alpha_3,\beta_3,\gamma_3)$ depend on $(\alpha_2,\beta_2,\gamma_2)$ in the exact same way as the latter depends on $(\Lambda_{1+},\Lambda_{1c},\Lambda_{1-})$. If we were to compose a fourth operator, the new coefficients would still be the same functions of the previous ones, by the same reasoning presented in Eqs. \eqref{eq:newBCH2} and \eqref{eq:newBCH3}. This pattern allows us to build recurrence relations for the coefficients of the $G$ operator resulting from the composition of $N$ group elements according to
\begin{align}
     G(\boldsymbol{\Lambda}_{N})\cdots G(\boldsymbol{\Lambda}_{1})=e^{\alpha_{N} T_{+}}e^{\ln(\beta_{N}) T_{c}}e^{\gamma_{N} T_{-}} \,,
\label{eq:Ncomp}
\end{align}
and
\begin{align}
    \alpha_{N}&=\Lambda_{N+}+\frac{\alpha_{(N-1)}(\Lambda_{Nc})^\delta}{1-\epsilon\delta\alpha_{(N-1)}\Lambda_{N-}} ,\nonumber \\ \beta_N&=\frac{\beta_{(N-1)}\Lambda_{Nc}}{(1-\epsilon\delta\alpha_{(N-1)}\Lambda_{N-})^\frac{2}{\delta}},  \label{eq:alphabetagamma}\\
     \gamma_N &=\gamma_{(N-1)}+\frac{\Lambda_{N-}(\beta_{(N-1)})^\delta}{1-\epsilon\delta\alpha_{(N-1)}\Lambda_{N-}}\,, \nonumber
\end{align}
with $\alpha_1=\Lambda_{1+}\,$, $\beta_1=\Lambda_{1c}$ and $\gamma_1=\Lambda_{1-}\,$. These results are the new BCH-like relations \cite{NewBCH}. An interesting fact is that the $\alpha_N$ coefficients, which are independent of $\gamma_N$ and $\beta_N$, can be written in the elegant form 
\small
\begin{equation}
\alpha_{j}=\Lambda_{j+}-\cfrac{(\Lambda_{jc})^{\delta}}{\epsilon\delta\Lambda_{j-}-\cfrac{1}{\Lambda_{(j-1)+}-
\cfrac{(\Lambda_{(j-1)c})^{\delta}}
{\epsilon\delta\Lambda_{(j-1)-} \, -\cfrac{1}{\ldots -\cfrac{1}{\Lambda_{1+}}}}}} \, . \\
\label{eq:alpharecursive}
\end{equation}
\normalsize
This expression is a generalized continued fraction (GCF), a mathematical object which appears in several 
areas, such as complex analysis and number theory \cite{khinchin1997continued}. 
Its recursive structure is particularly suitable for a numerical implementation.

\subsection{Application to time-dependent Hamiltonians} 
\label{subsection:NewBCH}

In this section, we use the above results to solve the equations of motion of quantum systems with time-dependent Hamiltonians, 
in the special case when the Hamiltonian is a linear combination of the generators of the Lie algebras presented in Table I.
Consider the time-dependent Hamiltonian given by
\begin{equation}
    H(t)=\eta_+(t) T_++\eta_c(t) T_c+\eta_-(t) T_- \;.
\end{equation}
The coefficients $\boldsymbol{\eta}(t)=(\eta_+,\,\eta_c,\,\eta_-)$ are complex functions of time, and they carry all the time dependence of the Hamiltonian, since the generators of the Lie algebras ($ T$ operators) are assumed to be time-independent. As for any closed quantum system, the time evolution of the density operator between two instants $t_0$ and $t$
is dictated by the Schr\"odinger equation, which can be cast in the form
\begin{equation}
    \rho (t)= U(t,t_0)\,\rho (t_0)\, U^\dagger(t,t_0),
\end{equation}
where $U$ is the time evolution operator (TEO).
In general, the solution for the TEO of time-dependent systems is given by the Dyson series, which can be very difficult to treat mathematically. 
However, whenever the Hamiltonian is a linear combination of generators of a finite Lie algebra, the Wei-Norman theory allows us to find the TEO by solving a set of coupled differential equations \cite{Wei_1963, Wei_1964}. Here, we use a \textit{time-splitting} approach, allowing us to use the algebraic methods previously explained to directly calculate the TEO  without the necessity of solving any (nonlinear) differential equation.

Let us split the time interval of the evolution $t_{0} \leq t \leq t'$ in discrete steps of size $\tau_s$. Then we can write the composition property of the TEO  in the following way
\begin{equation}
 U(t',t_0)= U(t_{N},t_{N-1}) U(t_{N-1},t_{N-2})\cdots U(t_{1},t_0)\, , 
 \label{eq:compoU}
\end{equation}
where $t_j=t_0 + j\tau_s$, and $\{j\in\mathbb{N} \mid 1\leq j\leq N\}$. If we make $\tau_s\rightarrow 0$, $N\rightarrow \infty\,$, keeping $t'-t_0=N\tau_s$, we can treat the Hamiltonian as constant in each infinitesimal time interval. For a constant Hamiltonian, the TEO is trivial and the composition can be written as
\begin{equation}
U(t,t_0)  = \lim_{\substack{N\rightarrow\infty\\ N\tau_s = t}}\,
e^{-\frac{i}{\hbar} H_N\tau_s}\cdots e^{-\frac{i}{\hbar} H_1\tau_s} \, ,
\label{eq:TEOin}
\end{equation}
where we choose without loss of generality the Hamiltonian value in the interval as $H_j\equiv H(t=t_{0}+j\tau_s)\,$. 
Note that in the above equation we write $t$ instead of $t'$ since this notation is no longer needed. 
For numerical implementation of this method, we choose the time step $\tau_s$ to be much smaller than the typical timescale of the variation of the Hamiltonian parameters $\boldsymbol{\eta}(t)\,$. In this case, we can safely assume these parameters to be constant by parts in each one of the time steps. 

It will be useful to write the stepwise time evolution in terms of the Hamiltonian parameters,
\begin{align}
\boldsymbol{\eta}(t)=\left\{
\begin{array}{ccc}
\boldsymbol{\eta}_{0}\, , \: & \mbox{for} & t\leq t_0, \\
\boldsymbol{\eta}_{1}\, , \: & \mbox{for} & t_0<t\leq\tau_s, \\
 \vdots & & \vdots \\
\boldsymbol{\eta}_{j}\, , \: &\mbox{for} & (j-1)\tau_s<t\leq j\tau_s, \\
 \vdots & & \vdots \\
\boldsymbol{\eta}_{N}\, , \: & \mbox{for} & (N-1)\tau_s<t\leq N\tau_s \, , \\
\end{array} \right.
\label{eq:discrete}
\end{align}
with $\boldsymbol{\eta}_j=(\eta_{j+},\,\eta_{jc},\,\eta_{j-})$. In accordance with the choice made for the Hamiltonian, we consider  $\boldsymbol{\eta}_j:=\boldsymbol{\eta}(t=t_{0}+j\tau_s)$. 
In this way, the constant Hamiltonian and the corresponding TEO 
within each time interval can be written as
\begin{eqnarray}
 H_j &=& \eta_{j+}\, T_++\eta_{jc}\, T_c+\eta_{j-}\, T_-\;,
\nonumber\\
 U_{j} &=& e^{\lambda_{j+} T_++\lambda_{jc} T_c+\lambda_{j-} T_-}\;,
\label{eq:TEOj}
\end{eqnarray}
where $\boldsymbol{\lambda}_j=-\frac{i}{\hbar}\tau_s\boldsymbol{\eta}_j\,$ and $U_j = U(t=t_{0}+j\tau_s)\,$. 

Note that the stepwise composition of the TEO, given in Eq. \eqref{eq:TEOin}, is similar to the composition shown in Eq. \eqref{eq:Ncomp}. Moreover, each step $U_j$ has the same form as the $G$ operators presented in Eq. \eqref{eq:Goperator1}. We can, therefore, apply the aforementioned factorization procedure to get
\begin{equation}
     U_{j}=e^{\Lambda_{j+} T_+}e^{\ln{\Lambda_{jc}} T_c}e^{\Lambda_{j-} T_-},
\end{equation}
where the relations between $\boldsymbol{\Lambda}$ and $\boldsymbol{\lambda}$ were given in Eqs. (\ref{eq:biglambdas}) and (\ref{eq:nu}). Now we insert the factorized form of the TEO for each time step 
in Eq. (\ref{eq:TEOin}) to obtain
\begin{align}
     &U(t,t_0)= U_N\, U_{(N-1)}\cdots U_2\, U_1 =\\
    &\{e^{\Lambda_{N+} T_{+}}e^{\ln(\Lambda_{Nc}) T_{c}}e^{\Lambda_{N-} T_{-}}\}
\cdots
\{e^{\Lambda_{1+} T_{+}}e^{\ln(\Lambda_{1c}) T_{c}}e^{\Lambda_{1-} T_{-}}\}\,.
\nonumber
\end{align}
We have shown that the TEO can be written as a sequence of products of $ G $ operators in the factorized form, one for each infinitesimal time step. This is exactly the same situation we found when we discussed the composition rule for elements of the given Lie groups. We can, therefore, use Eq. \eqref{eq:Ncomp} 
to rewrite the TEO in the final form,
\begin{equation}
     U(t,t_0)=e^{\alpha_{N} T_{+}}e^{\ln(\beta_{N}) T_{c}}e^{\gamma_{N} T_{-}},
\end{equation}
where the coefficients $(\alpha_N\,,\beta_N\,,\gamma_N$) are given in Eqs.  (\ref{eq:alphabetagamma}).
The recursive relations for the coefficients mentioned above are very well suited for numerical implementations, and they have been used to describe the dynamics of a harmonic oscillator with a time-dependent frequency \cite{efficientsolutionTDHO, DMT-JPHYSB-2021}. 

\subsection{The quantum quench limit}
\label{subsec:quench}

There is an interesting scenario to consider within this framework that will be useful for our analysis in Sec. \ref{sec:Performance_Optimization}, and that is the one where the Hamiltonian changes instantaneously from an initial $H_i$ to another value $H_f$ at a given instant $t_q\,$, with corresponding parameters $\boldsymbol{\eta}_i$ and $\boldsymbol{\eta}_f$. In the literature this type of protocol is refereed to as a ``\textit{quench}" in the Hamiltonian, and it can be described by a single iteration of the method, or equivalently, as a time-splitting described in Eq. \eqref{eq:discrete} for $N=2$. In this case, the Hamiltonian can be written as 
\begin{eqnarray}
    H(t) &=& \Theta(t-t_q) H_f + \left[1-\Theta(t-t_q)\right] H_i\,,
    \label{eq:Hquench}
\end{eqnarray}
where $\Theta$ is the usual Heaviside step function. Therefore, the time evolution operator, $U_q$, is given by
\begin{eqnarray}
    U_q (t,t_0) &=& e^{-\frac{i}{\hbar}H_f (t-t_q)}\, e^{-\frac{i}{\hbar}H_i (t_q-t_0)} \nonumber\\
    &=& e^{\alpha_{q} T_{+}}e^{\ln(\beta_{q}) T_{c}}e^{\gamma_{q} T_{-}}\,,
    \label{eq:quench}
\end{eqnarray}
%
as it is just a composition of two group elements, as in Eq. \eqref{eq:compo}.  
The ($\alpha_q,\beta_q,\gamma_q$) coefficients can thus be easily obtained in terms of the parameters of the initial and final Hamiltonians, $\boldsymbol{\eta}_i$ and $\boldsymbol{\eta}_f$, using Eqs. \eqref{eq:biglambdas}, \eqref{eq:nu} and \eqref{eq:newBCH1}.

\section{The isotropic Heisenberg model}
\label{section:model}

In this section we will use the algebraic methods described above to analyze an Otto cycle with a working substance composed of two interacting spins, or qubits. The spins are driven by a global interaction with a time-dependent external magnetic field, which can vary in direction and intensity. We also consider an interaction between the two qubits analogous to the one in the Heisenberg XXX (or isotropic) Hamiltonian for two spins \cite{Statistical_Many_Body}. An Otto cycle with this interaction model was presented in Refs. \cite{2qbitOttoengine1} and \cite{OttoCycleUFF}. However, these previous works considered the magnetic field fixed in one direction, which, in the isotropic case, causes the Hamiltonian to commute at different times. 

Let us start by describing the working substance and its corresponding Hamiltonian. Since we are dealing with a system of interacting spins, the total Hamiltonian of the working substance will be composed of a term associated with the external driving of the magnetic field, as well as an internal term, related to the coupling. We will discuss the properties of these two terms, and  use our results to compute the TEO for this system.

 \subsection{The external Hamiltonian}
 \label{sec:externalhamiltonian}
 
 We consider an external magnetic field that couples to the total magnetic moment of the two spins. This magnetic moment is proportional to the the global angular momentum operator: $\mathbf{S}=\mathbf{S}_1+\mathbf{S}_2$, and the magnetic moment is $\mathbf{\mu}=g_m \mathbf{S}\,$, where $g_m$ is the gyromagnetic constant. The Hilbert space of the total angular momentum operator of this system can be divided into the \textit{singlet} and \textit{triplet} subspaces \cite{Cohen-Tannoudji}. All three components of the total angular momentum operator, $\mathbf{S}=(S_x,S_y,S_z)$, are block diagonal in these subspaces. They are defined in terms of the individual angular momenta by
 \begin{eqnarray}
     S_i &=& \frac{\hbar}{2}\left(\Sigma_{i1} + \Sigma_{i2}\right)
     \equiv\frac{\hbar}{2}\,\Sigma_i\;,
     \nonumber\\
     \Sigma_{i1} &=& \sigma_{i1}\otimes\mathbb{1}_2\;,
     \quad
     \Sigma_{i2} = \mathbb{1}_1\otimes\sigma_{i2}\;.
     \label{eq:globalangularmomentum}
 \end{eqnarray}
where $\sigma_{ik}$ is the $i$-th Pauli matrix acting on the $k$-th spin subspace ($i=x,y,z$ and $k=1,2$).

We define $H_{ext}$ as the part of the Hamiltonian associated with the interaction with the magnetic field, 
namely
\begin{eqnarray}
H_{ext}(t) &=& -\mathbf{\mu}\cdot \mathbf{B}(t) \nonumber\\ 
&=& X(t)\,\Sigma_x + Y(t)\,\Sigma_y +Z(t)\,\Sigma_z\,,
\label{eq:ExtH} 
\end{eqnarray}
where $(X,Y,Z)\equiv -g_m\frac{\hbar}{2}(B_x,B_y,B_z)$. Note that for a time varying magnetic field, with 
variable amplitude and direction, $[H_{ext}(t),H_{ext}(t^\prime)]\neq 0\,$. However, this Hamiltonian is 
block diagonal in the singlet and triplet subspaces, and we can define the Rabi frequency as
\begin{eqnarray}
    \omega(t) &=& \frac{2}{\hbar}\sqrt{X^2(t)+Y^2(t)+Z^2(t)}\;. \label{eq:energyeigenvalue2qbit}
\end{eqnarray}
With this definition the eigenvalues of the external Hamiltonian become $+\hbar\omega(t)$,$\,0$ and $-\hbar\omega(t)$, with a double degeneracy in the null eigenvalue.

It is useful to define the global ladder operators as
\begin{eqnarray}
    \Sigma_\pm &=& \Sigma_x \pm i\,\Sigma_y\;, 
    \label{eq:ladderoperators}
\end{eqnarray}
and write the external Hamiltonian as
\begin{eqnarray}
    H_{ext}(t) &=& \eta_+(t)\, T_+ + \eta_c(t)\, T_c + \eta_-(t)\, T_- \;, \nonumber
\end{eqnarray}
where
\begin{eqnarray}
\eta_\pm(t)  &=&  X(t)\mp i Y(t)\,, \quad \eta_c(t)  = 2 Z(t)\,,\\
T_\pm  &=&  \frac{\Sigma_{\pm}}{2}\,, \quad T_c = \frac{\Sigma_z}{2}\,. 
\end{eqnarray}
Note that $\,T_+,T_c$ and $T_-$ satisfy the commutation relations of the $\mathfrak{su}(2)$ algebra. 
Using the methods discussed Sec. \ref{sec:algebra}, we conclude that the time evolution operator 
generated by an arbitrary driving of the magnetic field can be cast in the form
\begin{eqnarray}
    U_{ext} &=& e^{\alpha\, T_{+}}e^{\ln{\beta}\, T_c}e^{\gamma\, T_{-}}\;,
    \label{eq:Uext}
\end{eqnarray}
where the parameters $(\alpha,\beta,\gamma)$ can be calculated numerically, as described in Sec. \ref{sec:algebra}.

\subsection{The interaction Hamiltonian} 
\label{sec:interactionhamiltonian}

For systems of interacting spins, there are numerous  models that describe a variety of situations, and the Heisenberg model for spin lattices is a well known example  \cite{Statistical_Many_Body}. Following Refs. \cite{2qbitOttoengine1}, \cite{2qbitOttoengine2} and \cite{2qbitOttoengine3}, we analyze the 1D isotropic Heisenberg model for the interaction. The Hamiltonian that describes the interaction between the two qubits is given by
\begin{eqnarray}
    H_{int} &=& \frac{8J}{\hbar^2}\, \mathbf{S}_{1}\cdot \mathbf{S}_2 - 2J\;,
    \nonumber\\
            &=& \frac{4J}{\hbar^2}\,\left(S^2 - S_1^2 - S_2^2\right) - 2J\;,
    \label{Hint}
\end{eqnarray}
where $J$ is the coupling parameter and the zero energy level was shifted to simplify its eigenvalues, which can be easily shown to be $\{-8J,0,0,0\}$. Note that the interaction Hamiltonian, $H_{int}\,$, is null in its degenerate subspace. The sign of the coupling parameter determines if we are in the antiferromagnetic ($J<0)$, or in the ferromagnetic ($J>0$) regime. We restrict ourselves to the antiferromagnetic case, which is more interesting since it is the only one that can exhibit quantum entanglement, as proven in Ref. \cite{2qbitOttoengine2}. This interaction is isotropic in the sense that the coupling parameter is the same for all three directions. For this reason, even if the coupling changes in time, the internal Hamiltonian will commute with itself at different times, and therefore the time evolution operator is trivial. Assuming a constant coupling parameter during a time interval $\Delta t=t-t_0\,$, we have
\begin{equation}
    U_{int}=\exp\left(-\frac{i}{\hbar} H_{int} \Delta t\right)\;. 
    \label{eq:Uint}
\end{equation}
We can see that the interaction Hamiltonian commutes with the components of $\mathbf{S}$, which implies
\begin{equation}
    [H_{ext}(t),H_{int}]=0\quad \forall t \;, 
    \label{eq:HextHint}
\end{equation}
which is a direct consequence of the interaction isotropy. 

The complete Hamiltonian we consider is the sum of the external and internal parts, namely,
\begin{eqnarray}
     H(t) &=& H_{ext}(t)+H_{int} 
     \nonumber\\
     &=& \frac{1}{2}\,\boldsymbol{\eta}(t) \cdot \boldsymbol{\Sigma} 
     + J\left( \Sigma^{2} - 8 \right)\,.
     \label{eq:globalhamiltonian} 
\end{eqnarray}
%
We can easily obtain the eigenvalues of $H(t)$:
\begin{eqnarray}
    E_1(t) &=& -8J\,, \quad E_2(t) = \hbar \omega(t)\,,
    \nonumber \\
    E_3(t) &=& 0\,, \qquad E_4(t) = -\hbar\omega(t)\,.
    \label{eq:energyeigenvalues} 
\end{eqnarray}
Differently from the case of a single qubit, the energy eigenvalues depend on the coupling constant as well as the Rabi frequency.

\subsection{The time evolution operator}
\label{sec:TEO}

Since the external and internal Hamiltonians 
commute, we can obtain the time evolution operator for the total Hamiltonian by composing $U_{ext}$ and $U_{int}$, given in Eqs. (\ref{eq:Uext}) and (\ref{eq:Uint}), respectively. The resulting TEO is
\begin{equation}
    U=U_{ext} \, U_{int}\;.
    \label{eq:Ucomplete}
\end{equation}
Moreover, it is possible to factorize each step of the evolution described in Eq. \eqref{eq:compoU}, 
and isolate the interaction contribution to the total evolution, allowing us to use the algebraic methods described 
in Sec. \ref{sec:algebra} for the calculation of $U_{ext}\,$. In mathematical terms, we can write
\begin{eqnarray}
 U(t,t_0) &=& e^{-iH_{int} (t-t_0)}\,U_{ext}(t,t_0)\, , 
 \nonumber\\
 U_{ext}(t,t_0) &=& U_{ext}(t,t_{N-1})\cdots U_{ext}(t_{1},t_0)
   \label{eq:compoUext}\\
 &=& e^{\alpha\, T_{+}}e^{\ln{\beta}\, T_c}e^{\gamma\, T_{-}}\;.
   \nonumber
\end{eqnarray}
We can benefit from algebraic properties of the generators $\{T_{\pm},T_c\}$ to obtain the matrix 
representation of the TEO in the singlet-triplet basis. First, note that $T_{\pm}$ 
are proportional to the ladder operators $\Sigma_{\pm}\,$, which satisfy 
$\left(\Sigma_{\pm}\right)^{\,n} = 0$ for $n\geq 3\,$. 
This means that the exponential terms involving $T_{\pm}$ are limited to the first three terms in 
their Taylor series. Moreover, $T_c$ is proportional to $\Sigma_z\,$, which is diagonal, so its 
exponential can be trivially determined. By combining these properties, we arrive at a rather simple 
matrix form for the total time evolution operator
\begin{eqnarray}
\!\!\!\!\!\!\!\!\!\!\!\!\!\!U(t,t_0) \!=\!\!  
\begin{bmatrix}
 e^{8 i J \Delta t /\hbar} & 0 & 0 & 0 \\
 0 & \frac{(\alpha  \gamma +\beta )^2}{\beta } & \frac{\sqrt{2} \alpha  (\alpha  \gamma +\beta )}{\beta } & \frac{\alpha ^2}{\beta } \\
 0 & \frac{\sqrt{2} \gamma  (\alpha  \gamma +\beta )}{\beta } & \frac{2 \alpha  \gamma }{\beta }+1 & \frac{\sqrt{2} \alpha }{\beta } \\
 0 & \frac{\gamma ^2}{\beta } & \frac{\sqrt{2} \gamma }{\beta } & \frac{1}{\beta } \\
\end{bmatrix}\!.
 \label{eq:Umatrix}
\end{eqnarray}
Note that the TEO is block diagonal in the global basis, which means that the singlet state
is stationary. Therefore transitions between the singlet and the other states will not occur in this model. 
The coefficients $(\alpha,\beta,\gamma)$ can be calculated numerically by the methods presented in Sec.
\ref{sec:algebra} for an arbitrary protocol of the magnetic field.

\subsection{The transition probability matrix}
\label{sec:pmatrix}

We consider a generic sweep of the magnetic field that takes the Hamiltonian given in Eq. \eqref{eq:globalhamiltonian} from $H_1\equiv H(t_1)$ to $H_2\equiv H(t_2)$. The instantaneous eigenvalues and eigenstates are given by
\begin{align}
    H_r \ket{E_m^{(r)}}=E_m^{(r)} \ket{E_m^{(r)}}, \:\:\:\:\: m=1,2,3,4,\:\:
    r=1,2,\label{eq:instantaneouseigenstates}
\end{align}
with
\begin{align}
    E_1^{(r)}&=-8J\,, && E_2^{(r)}=\hbar\omega_{r}\,, \nonumber\\ 
    E_3^{(r)}&=0\,, && E_4^{(r)}=-\hbar\omega_{r}\,. \label{eq:eigenvaluest1t2}
\end{align}
We have defined $\omega_r\equiv\omega(t=t_r)$, with $\omega(t)$ given by Eq. \eqref{eq:energyeigenvalue2qbit}. Note that the energy eigenvalues depend on both the coupling parameter and the Rabi frequency.

The work probability distribution and all the relevant thermodynamic quantities 
can be expressed in terms of the elements of the transition probability matrix, which is given by
\begin{equation}
    p_{i|j}=\abs{\bra{E_i^{(2)}}U\ket{E_j^{(1)}}}^2\,.
    \label{eq:transitionprobabilities}
\end{equation}
This matrix is symmetric and doubly stochastic, so
\begin{eqnarray}
    p_{i|j} &=& p_{j|i}\;,
    \nonumber\\
    \sum_j p_{i|j} &=& \sum_i p_{i|j}=1\;.
\label{eq:doublestochasticity}
\end{eqnarray}
Using these properties, we can solve a simple linear system to write all the 16 entries of $p_{i|j}$ in terms of only the diagonal elements, which are the persistence probabilities of the instantaneous eigenstates. We can simplify this further by using the fact that the TEO is block-diagonal, which means that no transitions occur between the singlet and triplet states. Therefore, the probability of persistence in the singlet state must be unity, and we can describe the elements of $p_{i|j}$ with only three parameters, namely, $P\equiv p_{2|2}$, $P'\equiv p_{3|3}$ and $P''\equiv p_{4|4}\,$. The transition matrix can therefore be written as
\begin{align}
    p_{i|j}=
\begin{bmatrix}
 1 & 0 & 0 & 0 \\
 0 & P & Q & Q' \\
 0 & Q & P' & Q'' \\
 0 & Q' & Q'' & P'' \\
\end{bmatrix} ,
\label{eq:pmatrix}
\end{align}
where
\begin{eqnarray}
    Q &=& \frac{1}{2} \left(-P'+ P''- P + 1\right)\,,
    \nonumber\\
    Q' &=& \frac{1}{2} \left(P'- P''- P + 1\right)\,,
    \nonumber\\
    Q'' &=& \frac{1}{2} \left(-P'- P''+ P + 1\right)\,.
    \nonumber\\
\label{eq:offdiagonal}
\end{eqnarray}

The three parameters ($P,P',P''$), that we call henceforth as $P-$parameters, are the persistence probabilities on the highest energy state, the zero energy state, and the ground state of the triplet, respectively. Note that these parameters are \textbf{not} independent, being jointly determined by the unitary evolution produced by the sweep in the external magnetic field. They are convenient because they help us to measure the degree of adiabaticity of the protocol, as they all approach unity in the adiabatic limit.
With the initial and final energy eigenvalues at hand, as well as the elements of the transition probability matrix, we have all the information needed for computing the desired thermodynamic quantities.

\section{Two qubits Otto cycle}
\label{section:twoqbitottocycle}

Now that we have described the working substance, let us investigate its behavior when operated in the same Otto cycle described in Ref. \cite{Campisi1}. The protocol is composed by a sequence of operations (strokes) interpolating four points of the cycle as follows:

\textit{Point A:} 
The substance is initially in a thermal state, at inverse temperature $\beta_h\,$, described by the density operator 
$\rho_A=e^{-\beta_h H_1}/Z_1\,$, which gives
\begin{align}
    E_A &=\Tr(\rho_A H_1) \nonumber\\
    &=-\frac{8 J e^{8 J \beta _h}}{Z_1}-\frac{2 \hbar \omega _1 \sinh \left(\beta _h\hbar \omega _1 \right)}{Z_1}\,, \label{eq:EA}
\end{align}
where $Z_1$ is the partition function given by
\begin{align}
Z_1=1+e^{8 J \beta _h}+e^{-\beta_h \hbar \omega_1}+e^{\beta _h \hbar \omega _1 }\,.
\label{partitionfunction}
\end{align}

\textit{Point B:} The system is subjected to the work protocol corresponding to a change in the external magnetic field, and the density operator evolves according to $\rho_B=U\rho_A U^\dagger$, while the Hamiltonian changes from $H_1$ to $H_2$. The average energy at this point depends only on the eigenvalues of the Hamiltonian at the initial and final instants of the transformation and the elements of the transition probability matrix given in Eq. \eqref{eq:pmatrix}. Therefore, we have 
\begin{align}
     &E_B = \Tr(\rho_B H_2 )
     \nonumber\\
     &=\frac{1}{Z_1} \sum_{i,j} e^{-\beta_h E_j^{(1)}}  E_{i}^{(2)} \abs{ \bra{E_i^{(2)}} U \ket{E_j^{(1)}} }^2
     \nonumber \\
     &=-\frac{8 J e^{8 J \beta _h}}{Z_1}-\frac{2 \hbar \omega _2 \sinh \left(\beta _h \hbar \omega _1 \right)}{Z_1}
     \label{eq:EBtwoqbit}\\
     &+\frac{\hbar\omega_2}{Z_1}\{f_1 (1\!-\!P) - f_2(1\!-\!P') + f_3 (1\!-\!P'')\}\,,
     \nonumber
\end{align}
where we defined
\begin{align}
    f_1&=1+\frac{e^{\beta_h \hbar\omega_1}}{2}-\frac{3e^{-\beta_h \hbar\omega_1}}{2}\,, \nonumber\\
    f_2&=\sinh{(\beta_h \hbar \omega_1)}\,, \nonumber\\
    f_3&=\frac{3e^{\beta_h \hbar\omega_1}}{2}-\frac{e^{-\beta_h \hbar\omega_1}}{2}-1\,. \label{eq:ffunctions}
\end{align}
The above are all positive functions for $\beta_h\hbar\omega_1>0\,$. 

It is instructive to inspect the behavior of the average energy in the adiabatic regime, where $(P,P',P'')\rightarrow(1,1,1)$. 
In this limit, the term in square brackets in Eq. \eqref{eq:EBtwoqbit} vanishes, and we trivially obtain
\begin{align}
    E_B^{(ad)}=-\frac{8 J e^{8 J \beta _h}}{Z_1}-\frac{2 \hbar \omega _2 \sinh \left(\beta _h \hbar \omega _1 \right)}{Z_1}\,.
\end{align}
Here, the term proportional to the coupling constant $J$ is equal to the one appearing in $E_A\,$, 
given in Eq. \eqref{eq:EA}. Moreover, if we make $J=0\,$, the two qubits become independent and we 
recover the result for a single qubit engine in the adiabatic regime, 
$E_B^{(ad)} = E_A \omega_2/\omega_1\,$, as can be easily checked \cite{Campisi1}.

\textit{Point C:} The system thermalizes with the cold source, so that $\rho_C=e^{-\beta_c H_2}/Z_2$ and
\begin{align}
    E_C &=\Tr(\rho_C H_2) \nonumber\\
    &=-\frac{8 J e^{8 J \beta _c}}{Z_2}-\frac{2 \hbar \omega _2 \sinh \left(\beta _c \hbar \omega _2 \right)}{Z_2}\,, \label{eq:EC}
\end{align}
where $Z_2=1+e^{8 J \beta _c}+e^{-\beta_c \hbar \omega_2}+e^{\beta _c \hbar \omega _2 }\,$.

\textit{Point D:} The magnetic field changes to its initial value and the Hamiltonian changes, in the backward protocol, from $H_2$ to $H_1$ with $\rho_D=U\rho_C U^\dagger$.
As it can be easily demonstrated, the matrix elements of the transition probability 
matrix are the same as in the first work protocol. We then have
\begin{align}
     &E_D = \Tr(\rho_D H_1 ) \nonumber\\
     &=\frac{1}{Z_2} \sum_{i,j} e^{-\beta_c E_j^{(2)}}  E_{i}^{(1)} 
     \abs{ \bra{E_i^{(1)}} \Tilde{U}_\tau \ket{E_j^{(2)}}}^2 
     \nonumber \\
     &=-\frac{8 J e^{8 J \beta _c}}{Z_2}-\frac{2 \hbar \omega _1 \sinh \left(\beta _c \hbar \omega _2 \right)}{Z_2} \label{eq:ED}\\
     &+\frac{\hbar\omega_1}{Z_2}\{g_1 (1\!-\!P) - g_2(1\!-\!P') + g_3 (1\!-\!P'')\}\,,
     \nonumber
\end{align}
with
\begin{align}
    g_1&=1+\frac{e^{\beta_c \hbar\omega_2}}{2}-\frac{3e^{-\beta_c \hbar\omega_2}}{2}\,,
    \nonumber\\
    g_2&=\sinh{(\beta_c \hbar \omega_2)}\,,
    \nonumber\\
    g_3&=\frac{3e^{\beta_c \hbar\omega_2}}{2}-\frac{e^{-\beta_c \hbar\omega_2}}{2}-1\,.
    \label{eq:gfunctions}
\end{align}

\subsection{Work and heat}
\label{section:WQ}

Given the average energy at the four points of the cycle, we have all the information necessary to compute the quantities describing the energy exchange between the working substance and the environment, namely work and heat. 
The heat exchanged with the hot source is given by

\begin{equation}
    Q_h = E_A-E_D = Q_h^{(l)}+Q_h^{(ad)}+Q_h^{(f)}\,, 
    \label{eq:Qh}
\end{equation}
where the three contributions are
\begin{eqnarray}
   Q_h^{(l)} &=& 8J(\frac{e^{8J\beta_c}}{Z_2}-\frac{e^{8J\beta_h}}{Z_1})\,,
   \nonumber\\
   Q_h^{(ad)} &=& \hbar \omega _1\left(\frac{2  \sinh \left(\beta _c \hbar \omega _2 \right)}{Z_2}-\frac{2  \sinh \left(\beta _h\hbar \omega _1 \right)}{Z_1}\right)\,,
   \label{Q1} \\
    Q_h^{(f)} &=& \frac{\hbar\omega_1}{Z_2}\{-g_1 (1-P) +g_2(1-P')-g_3 (1-P'')\}\,.  
    \nonumber
\end{eqnarray}
The adiabatic terms, $Q_h^{(l)}$ and $Q_h^{(ad)}$ recover the results presented in \cite{2qbitOttoengine1}, where this same thermodynamic cycle was analyzed in the particular case of quantum adiabatic processes in the driving strokes, and a magnetic field with fixed direction.
The contribution $Q_h^{(l)}$ was interpreted as a heat leak between the two heat reservoirs. 
It depends on the coupling constant $J$ both explicitly and implicitly via the partition functions 
$Z_1$ and $Z_2\,$. This implicit dependence on $J$ is shared with the other terms as well. It was also 
shown in Ref. \cite{2qbitOttoengine1} that this term cannot be harnessed as useful work, but it can improve the efficiency of this model in comparison with the uncoupled case for the domain of parameters where it becomes negative, which makes the heat flow from the cold to the hot source. This enhancement in efficiency will be exploited in Sec. \ref{sec:Performance_Optimization}. We have also $Q_h^{(ad)}$, the part of the heat exchanged in adiabatic processes, which does not depend explicitly on the coupling.

The last term, $Q_h^{(f)}$, is a consequence of a nonadiabatic driving of the working substance. Since the functions $g_1$, $g_2$ and $g_3$ are all positive, this term is an increasing function of the parameters $P$ and $P''$, and decreasing in $P'$. This happens because $P'$ is the persistence probability of the instantaneous eigenstate with energy $E_3(t)=0$. Consequently, we see that populating states with static energy levels will disrupt the heat exchange, and also reduce the overall efficiency of the machine. In Refs. \cite{quantumfriction1} and \cite{quantumfriction2}, this additional heat is interpreted as the quantum analog of friction, and it arises from transitions induced in the working substance.

To compute the heat exchanged with the cold source, we point out that the expressions of $(E_A,E_B)$ and $(E_C,E_D)$ are symmetrical with respect to the change $\hbar\omega_1 \leftrightarrow \hbar\omega_2$ and $\beta_h\leftrightarrow\beta_c\,$. Therefore, the expressions for the exchanged heats also have this symmetry. We then have
\begin{equation}
    Q_c = E_C-E_B = Q_c^{(l)}+Q_c^{(ad)}+Q_c^{(f)} \,,
    \label{eq:Qc}\\
\end{equation}
with
\begin{align}
    Q_c^{(l)}&=8J(\frac{e^{8J\beta_h}}{Z_1}-\frac{e^{8J\beta_c}}{Z_2})\,,
    \nonumber\\
    Q_c^{(ad)}&=\hbar \omega _2\left(\frac{2  \sinh \left(\beta _h \hbar \omega _1 \right)}{Z_1}-\frac{2  \sinh \left(\beta _c \hbar \omega _2 \right)}{Z_2}\right)\,,
    \label{eq:Q2}\\
    Q_c^{(f)}&= \frac{\hbar\omega_2}{Z_1}\{-f_1 (1-P) +f_2(1-P')-f_3 (1-P'')\}\,. 
    \nonumber
\end{align}
As in the case of $Q_h$, the heat exchanged with the cold source is also an increasing function of the persistence probabilities $P$ and $P''$, and decreasing in $P'$, because $f_1$, $f_2$ and $f_3$ are positive. We can also see that the heat leak term is the negative of the one found for the heat exchange with the hot source, \textit{i.e.}, $Q_c^{(l)}=-Q_h^{(l)}$. Because of this, both terms cancel out in the total work of the cycle, given by
\begin{align}
    W&=Q_h+Q_c 
    \nonumber\\
    &=Q_h^{(ad)}+ Q_c^{(ad)}+Q_h^{(f)}+ Q_c^{(f)} 
    \label{eq:W}.
\end{align}
The additional work associated with friction is just the sum of the two heat terms that appear when the driving is not adiabatic, $W^{(f)}=Q_h^{(f)}+ Q_c^{(f)}$.

\section{Operation regimes}
\label{sec:Operation}

\begin{figure} \centering
\includegraphics[scale=0.3]{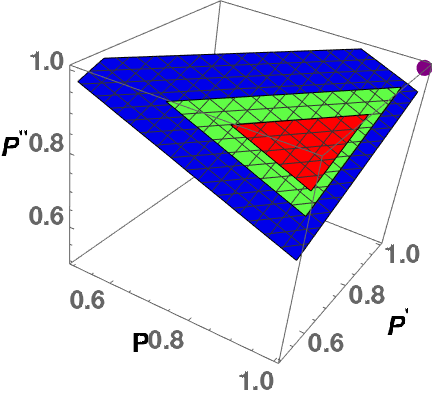} 
\includegraphics[scale=0.3]{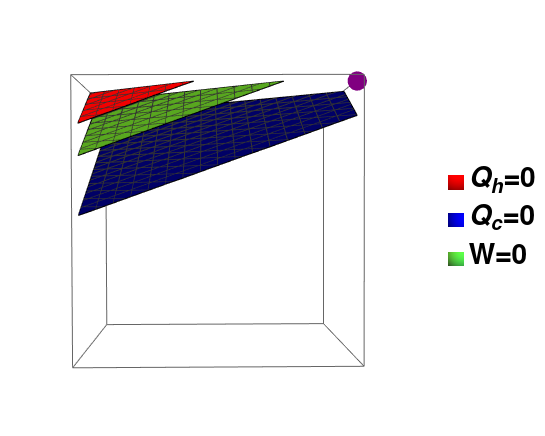} 
\caption{Zero-level planes of the exchanged heats and work. Here, we use the values $\hbar\omega_1/E_0=2\sqrt{5}\,$, $J/E_0=0.125\,$, $k_b T_h/E_0=2\,$, $T_c/T_h=0.5\,$, $\omega_2/\omega_1=1/\sqrt{5}\,$. The dot at point $(1,1,1)$ (top right corner) indicates the adiabatic regime. For better visualization, the probability domain has been reduced and we show two angles of the same plot.}
\label{fig:Jplanes2}
\end{figure}

The signs of the exchanged energies $Q_h$, $Q_c$, and $W$ will determine whether the machine will operate as an engine, a thermal accelerator, a refrigerator, or a heater. 
One of our aims is to understand how the adiabaticity of the protocol can influence the operation regimes of the machine. Note that the exchanged heats and work given in Eqs. \eqref{eq:Qh}, \eqref{eq:Qc}, and \eqref{eq:W}, are linear functions of the persistence probabilities $P$, $P'$, and $P''$, so the zero-level surfaces of these functions are expressions of the type $F(P,P',P'')=0\,$, and they represent planes in the domain of $(P,P',P'')$, contained within a cubic volume defined by $0\leq P,P',P'' \leq 1$ that we call the persistence probability space. These planes are shown in Fig. \ref{fig:Jplanes2}, where the probability domain has been reduced for better visualization. For a given point $(P,P',P'')$ in this space, the associated values of $Q_h\,$, $Q_c$, or $W$ will be positive if the point is above the associated plane, and negative if it is below. 
It is important to notice that the position of the planes is independent of the protocol. Moreover, a physical protocol associated with some driving of the magnetic field is represented by a point in this space. The point $(1,1,1)\,$, which describes the adiabatic limit, is represented by a dot in the top right corner of Fig. 1.

Since the work is the sum of the heats, the plane associated with $W=0$ is always between the other two planes. We have then two main distinct cases. In the first one, the plane associated with $Q_h=0$ is above the one corresponding to $Q_c=0$, and in the second one, the opposite holds true. Considering the constraints on the heats and work given by the first and second law of thermodynamics, only the heater and refrigerator are possible operation modes in the first case \cite{Campisi1}. By the same reasoning, the second case can admit only the engine, accelerator and heater operations.

\subsection{Protocol with a time-varying transverse field} \label{sec:protocol}
So far we have dealt with the system in a general manner regarding the possible work protocols, or unitary transformations, that can be applied to the working substance. All the information about the work protocol 
is encoded in the persistence probabilities, $(P,P',P'')$. Now we calculate these quantities explicitly, and 
analyze the results for a given protocol.

\begin{figure}[h] \centering
\includegraphics[scale=0.3]{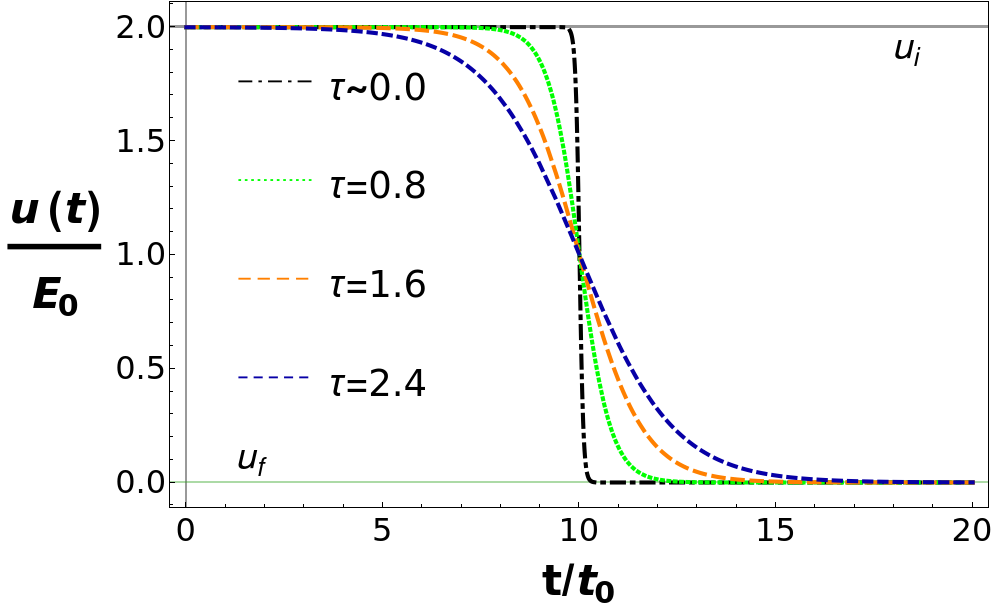} \caption{The control parameter associated with the external magnetic field in the $z$ direction, $u(t)/E_0$, as a function of time. 
We set the values $t_1/t_0=0$, $t_2/t_0=20$, $u_i=2E_0$ and $u_f=0$.
Note that for larger values of $\tau$, the time variation of the field becomes less abrupt.}
\label{fig:graphthg}
\end{figure}

Following the context of two spins driven by an external magnetic field presented in Sec. \ref{section:model}, we consider a protocol where the field is held constant in the $x$ direction, and varies in the $z$ direction as a hyperbolic tangent function. Therefore, in the expression for the external Hamiltonian, presented in Eq. \eqref{eq:ExtH}, we have
\begin{align}
    X(t)=\Delta\,, && Y(t)=0\,, && Z(t)=u(t)\,,
    \label{eq:control2qbits}
\end{align}
where $\Delta$ and $u$ have dimensions of energy and are proportional to the amplitude of the magnetic field in the $x$ and $z$ directions, respectively. We consider $\Delta$ to be constant and $u(t)$ given by
\begin{eqnarray}
    u(t) &=& \frac{(u_f-u_i)}{2} \tanh \left(\frac{t-t_q}{t_0\tau}\right) + \frac{u_f+u_i}{2}\,,
    \label{eq:protocolhypertange}
\end{eqnarray}
where $t_q$ was defined as the quench time in Sec. \ref{subsec:quench} and $t_q=\frac{t_1+t_2}{2}$, with $t_1$ and $t_2$ being, respectively, the initial and final evaluation instants defined in Sec. \ref{sec:pmatrix}, such that $u(t_1)=u_i$ and $u(t_2)=u_f$. 
For $X(t)$ and $Z(t)$ defined above, the energy levels of the working substance become
\begin{align}
    E_1^{(r)}&=-8J\,, && E_2^{(r)}=2\sqrt{\Delta^2+u_r^2}\,, \nonumber\\ 
    E_3^{(r)}&=0\,, && E_4^{(r)}=-2\sqrt{\Delta^2+u_r^2}\,, \label{eq:eigenvaluest1t2v2}
\end{align}
 with $r=i,f$. To show our results in terms of dimensionless quantities, we introduce a typical qubit energy scale $E_0$ and a timescale given by $t_0=h/E_0$, where $h$ is Planck's constant.
 In this section, for illustrative purposes, the asymptotic limits of $u(t)$ were chosen to be $u_i=2E_0$ and $u_f=0$, therefore the initial and final Bohr frequencies given by Eq. \eqref{eq:energyeigenvalue2qbit} are $\hbar\omega_1=2\sqrt{5}E_0$ and $\hbar\omega_2=2E_0\,$, which leads to $\omega_2/\omega_1=1/\sqrt{5}\,$. The nondimensional parameter $\tau$ is introduced to control the rate of change of the driving field of the protocol. Indeed, we can note from Fig. \ref{fig:graphthg} that as $\tau$ increases, the variation of the magnetic field takes a larger time interval to occur, and we approach the adiabatic limit as $\tau\rightarrow \infty$. Because of this, we call $\tau$ the adiabatic parameter.

With the expressions for the time-dependent Hamiltonian obtained in Eq. \eqref{eq:globalhamiltonian} and the resulting TEO given in Eq. \eqref{eq:Umatrix}, we numerically compute the persistence probabilities $(P,P',P'')$ 
for the given variation of the external magnetic field. 
For each possible trajectory determined by the parameters ($\Delta,u_i,u_f$,$\tau$,$J$) there will be a point in this space. To study how the adiabaticity of the work protocol can influence the operation regimes of the machine, we compute the persistence probabilities for different values of parameter $\tau$ and fixed values of the other parameters. Each value of $\tau$ will result in a triple $(P,P',P'')$ represented by a point in the space, and a sequence of values of $\tau$ will be represented by a parametric trajectory given by 
$\left[P(\tau),P'(\tau),P''(\tau)\right]\,$, which approaches $(1,1,1)$ in the adiabatic limit, 
\begin{equation}
    \lim_{\tau\rightarrow\infty} \left[P(\tau),P'(\tau),P''(\tau)\right] = (1,1,1)\,.
\end{equation}
When this parametric trajectory crosses a zero-level plane of the exchanged energies, we have a transition between two operation modes. 

\subsection{Operation transition}

\begin{figure}[h] \centering
\includegraphics[scale=0.31]{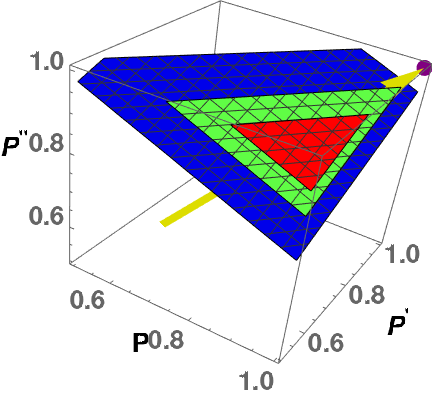}
\includegraphics[scale=0.33]{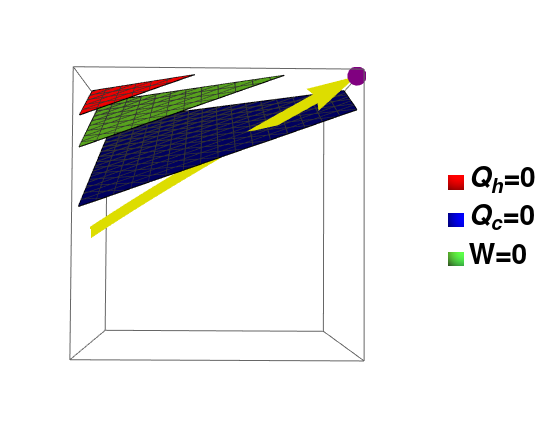}  
\caption{Transition between different operation regimes as the speed of the magnetic field variation is changed.
The yellow (solid) line represents the points associated with different values of the adiabatic parameter $\tau$\,. 
The zero-level planes are determined by the following values of the remaining parameters: $\hbar\omega_1/E_0=2\sqrt{5}$, $J/E_0=0.125\,$, $k_b\,T_h/E_0=2\,$, $T_c/T_h=0.5\,$, $\omega_2/\omega_1=1/\sqrt{5}\,$. As the sweep time $\tau$ increases, the machine changes the operation regime from heater (below the lower plane) to the refrigerator (above the lower plane). We show two angles of the same plot for better visualization of the points where the operation regime changes.} \label{fig:graphplanescurve1}
\end{figure}

In Fig. \ref{fig:graphplanescurve1}, we show how these transitions occur for some set of the system's parameters. 
We can see that as the adiabaticity of the protocol increases, the machine changes from a heater to a refrigerator, and therefore we can see that the operation mode of the machine depends directly on the timescale of the work protocol.

\begin{figure}[h] \centering
\includegraphics[scale=0.31]{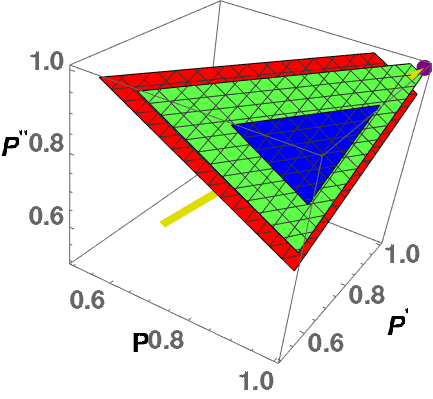}
\includegraphics[scale=0.33]{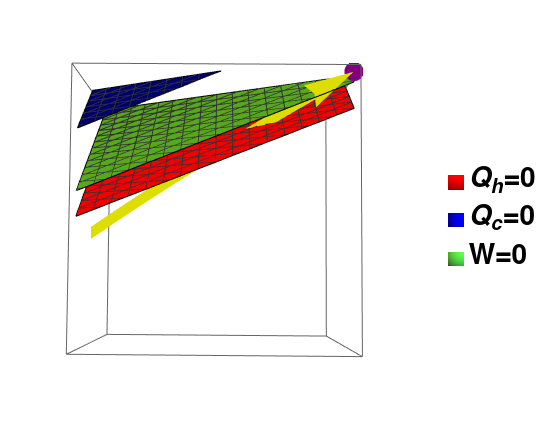}
\caption{Transition between different operation regimes as the speed of the magnetic field variation is changed.
The zero-level planes are determined by the following values of the parameters: $\hbar\omega_1/E_0=2\sqrt{5}$, $J/E_0=0.125$,$\,k_b\,T_h/E_0=4.7$,$\,\,T_c/T_h=0.375$,$\:\omega_2/\omega_1=1/\sqrt{5}$. The same parametric trajectory of Fig. \ref{fig:graphplanescurve1} is drawn. However, for the bath temperatures $T_c$ and $T_h$ considered here, 
the zero-level planes change and an engine operation becomes possible.
} \label{fig:graphplanescurve2}
\end{figure}

It is important to notice that the parametric trajectory (solid arrow) as a function of $\tau$ depends only on the characteristics of the coupling and the external magnetic field, and not on the temperatures of the hot and cold sources, $T_c$ and $T_h\,$. As can be seen in the expressions for the exchanged energies presented in Eqs. \eqref{eq:Qh}, \eqref{eq:Qc} and \eqref{eq:W}, these temperatures can modify the positions of the zero-level heat and work planes, and, therefore, change accessible  operation regions of the machine as a function of the persistence probabilities. In Fig. \ref{fig:graphplanescurve2} we show the same parametric trajectory as the previous case but with different values of $T_c$ and $T_h$. Note that, for this set of parameters, the operation mode changes from heater, to accelerator, and then to engine as the adiabaticity of the protocol ($\tau$) increases. Note also that the engine only becomes possible for protocols close to the adiabatic limit. %

\section{Friction and coupling under extreme operation regimes} \label{sec:Performance_Optimization}

\begin{figure}[h]  \centering
\includegraphics[scale=0.5]{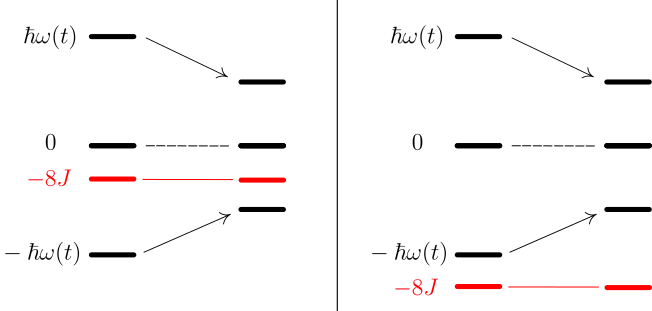}
\caption{Diagram of the energy levels of the working substance of the thermal machine in a compression stroke. In the  weak coupling regime (left), the ground state is the level $E_4=-\hbar\omega(t)$ that changes in time. Alternatively, in the strong coupling regime the ground state is the level $E_1=-8J$, which does not change in time. The time variation of the ground state of the working substance can drastically change the overall performance of the thermal machine, especially for small temperatures of the heat source, as discussed in the paper.}
\label{fig:leveldiagram}
\end{figure}

As we have mentioned before, the working substance has four nondegenerate energy levels, given in Eq. \eqref{eq:energyeigenvalues}. In the antiferromagnetic regime ($J>0$), the ground state of the system can be either $E_1=-8J$ or $E_2(t)=-\hbar\omega(t)$, depending on how the the coupling parameter $J$ compares with the energy associated to the magnetic field, $\hbar\omega(t)$. This level structure leads us to divide the analysis of the machine in two distinct regimes regarding the configuration of the ground state. The first, which we call the weak coupling regime ($8J<\hbar\omega(t))$, is the one where the level associated with the time-dependent intensity of the magnetic field remains as the ground state throughout the driving protocol. Conversely, the strong coupling regime ($8J>\hbar\omega(t)$) will happen when the level $E_1=-8J$ is the ground state. In Fig. \ref{fig:leveldiagram} we show a diagram illustrating these two regimes of the coupling parameter. 

\begin{figure}[h]  \centering
\includegraphics[scale=0.35]{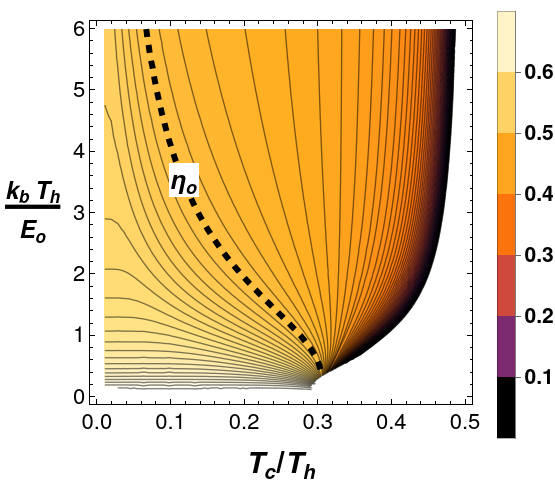}
\includegraphics[scale=0.35]{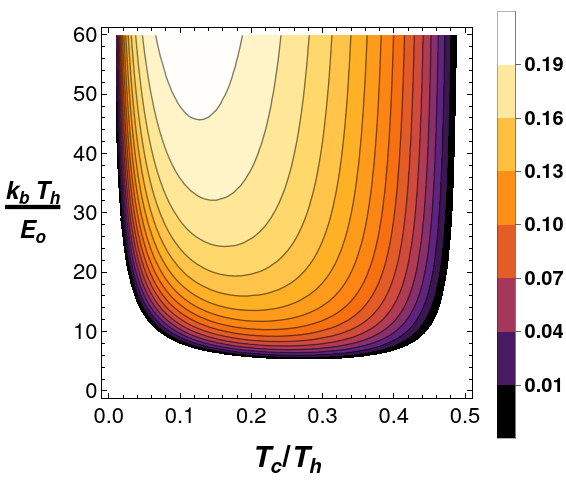} 
\caption{Efficiency of the engine as a function of the ratio between the bath temperatures and the temperature of the hot source in the regimes of weak and strong coupling. We have fixed the Rabi frequencies as $\hbar\omega_1/E_0=2$ and $\hbar\omega_2/E_0=1\,$, so the efficiency of the uncoupled model is $\eta_o=1-\omega_2/\omega_1=0.5\,$. Two values of the coupling parameter were considered. Top panel: In the weak coupling regime ($J/E_0=0.125$), the efficiency is maximum near the origin, and it surpasses the value of the uncoupled model for a wide range of temperatures. Bottom panel: For the strong coupling regime ($J/E_0=0.25$), the efficiency is much reduced in comparison with the case of weak coupling, and for small values of the bath temperatures the machine can even cease to operate as an engine.}
\label{fig:fig6}
\end{figure}

In this section, we aim to explore the differences in the operation of the machine in both these regimes. To address the matter of work extraction ($W>0$) we consider the engine as the operation regime, where the efficiency is given by $\eta=W/Q_h=1-Q_c/Q_h$. Initially, we show how 
these two regimes differ in the case of an adiabatic driving of the working substance. We then proceed to analyze the performance of the engine in finite-time, and we show how the effects of the  coupling can increase or decrease the energy losses due to quantum friction. 

First regarding the case of an adiabatic evolution, it is important to notice that, in this limit, the thermodynamic quantities such as heat and work will not depend on the driving protocol, but only on the initial and final values of the magnetic field. This can be seen directly from the expressions of $Q_h$, $Q_c$ and $W$, given in Eqs. \eqref{Q1}, \eqref{eq:Q2} and \eqref{eq:W}. For an adiabatic protocol, the persistence probabilities approach unity, and therefore the friction terms  $Q_h^{(f)}$ and $Q_c^{(f)}$ vanish. The remaining terms depend only on the bath temperatures, the coupling parameter, and the initial and final values of the Rabi frequency. This means that any protocol with the same initial and final Rabi frequencies will generate the same amounts of heat and work, provided that the timescale associated with the driving is much slower than the typical timescale of the working substance, given by the Bohr frequencies between the energy levels.

To illustrate how the two coupling regimes differ in the performance of the engine, we first analyze the adiabatic efficiency, which is the highest possible value of any protocol, since the energy losses caused by friction are absent in this case. In Refs.  \cite{2qbitOttoengine1} and \cite{OttoCycleUFF}, it is shown that the coupling between the spins of the working substance allows the machine to operate with an efficiency greater than the one of a system of uncoupled spins, which is $\eta_o = 1-\omega_2/\omega_1$. It is also shown that this enhancement is caused by the abnormal flux of heat directly between the two thermal reservoirs, that we defined as the heat leak terms, $Q_h^{(l)}=-Q_c^{(l)}$ 
[see Eqs. \eqref{Q1} and \eqref{eq:Q2}], that have a strong dependence on the coupling parameter $J$. 

In Fig. \ref{fig:fig6} we plot the efficiency as a function of the temperature of the hot source $T_h$ and the ratio between the temperatures $T_c/T_h$ for both coupling regimes. We highlight the level curve associated with the value of the uncoupled efficiency $\eta_o$ to identify the conditions for enhanced performance. We have also restricted the plot to the region where the machine operates as an engine, or equivalently, where $W>0$ and $\eta>0$. 

Comparing the efficiency in both regimes we can spot some striking differences. First, we can see that in the weak coupling regime the highest values of the efficiency are near the origin, when the bath temperatures are both small. Moreover, the efficiency surpasses $\eta_o$ for various values of the bath temperatures. For the strong coupling regime the behavior is considerably different. In this regime the efficiency is generally much smaller than the one of the weak coupling regime, and it does not surpass the uncoupled efficiency for any value of the bath temperatures. We also highlight that for the strong coupling regime, for given a fixed value of $T_h$ there is not just an upper, but also a lower limit for the values of $T_c$, and we get the maximum value of the efficiency only for high values of the bath temperatures ($k_b T_h >> E_0$).

The behavior described above has its physical origin in the energy level structure of the model considered, as pointed out in \cite{OttoCycleUFF}. In the strong coupling regime, the time-independent energy level $-8J$ represents a fixed ground state. As a consequence, in the thermalization stroke with the cold source, the population is heavily concentrated in a static energy level, while work extraction requires energy levels changing in time. This explains why the overall efficiency is reduced in the strong coupling regime. Moreover, this effect becomes more severe for smaller values of the bath temperatures, and it can even make the machine stop operating as an engine, as we can see in the behavior of the efficiency near the origin in Fig. \ref{fig:fig6}.

This behavior can be understood from the expressions given in Eqs. \eqref{Q1} and \eqref{eq:Q2} for the exchanged heat, from which one also obtains 
the work done 
and the efficiency $\eta=W/Q_h$ of the engine. We can see that these quantities depend on exponential functions of either $8J/k_b T$ or $\hbar\omega/k_b T\,$. For low temperatures, these factors become very large and the relevant thermodynamic quantities become very sensitive to the relative magnitude of the internal and external coupling terms.

\subsection{Finite-time performance for weak and strong coupling}

We now proceed to study the properties of this thermal machine outside of the adiabatic limit, where the working substance is driven in finite-time. We aim to describe the properties of the fraction of the exchanged heat that originates as a consequence of finite-time driving, \textit{i.e.}, the friction heat $Q^{(f)}=Q_h^{(f)}+Q_c^{(f)}$.

There is a protocol that can help us gain some intuition on the physics of the finite-time performance, the one where the function $u(t)$ behaves as a step function, meaning that the field in the $z$ direction is instantaneously switched between two values. This kind of driving, which was already considered at Sec. \ref{subsec:quench}, is usually called a ``\textit{quench}'' protocol \cite{CBPF1}. 

There are two main reasons for analyzing this type of protocol. First, it allows us to describe the dynamics of the working substance analytically, as the expressions for the permanence probabilities become very simple. Second, the quench is the limit of all protocols that share the same initial and final values of the field and the driving is infinitely fast compared with the typical timescales of the working substance, just like the adiabatic regime is the limit of all protocols with infinitely slow drivings. 

Consider a protocol as the one described in Eq. \eqref{eq:control2qbits}, where the function $u(t)$ behaves as a step function that falls to zero at some instant $t_q\,$.
The expression for $u(t)$ can be written as
\begin{equation}
 u(t) = \left\{
        \begin{array}{ll}
            \Delta\,, & \quad t \leq t_q \\
            \,0\,\,, & \quad t > t_q\;.
        \end{array}
    \right.
\end{equation}
We have chosen the value of the driving function before the quench to coincide with the constant value of the $x$ component of the magnetic field to simplify the notation and calculations, but this choice does not alter the physics we wish to describe.
Note that the step function corresponds to the limit $\tau\rightarrow 0$ of the protocol given in Eq. (\ref{eq:protocolhypertange}),
which reinforces the idea that the quench is the limit of any infinitely fast protocol.

Considering this driving function, the external Hamiltonian of the system becomes constant by parts. Before the quench, the Hamiltonian is $H_1=\Delta(\Sigma_x+\Sigma_z)$ [see Eq. \eqref{eq:globalangularmomentum}] with Rabi frequency $\hbar\omega_1=2\sqrt{2}\Delta$, and after the quench it is $H_2=\Delta\Sigma_x$ with $\hbar\omega_2=2\Delta$. 
Recall that, as we have discussed in Sec. \ref{subsection:NewBCH}, a quench protocol can be described by a single iteration in the algebraic method.

With the time evolution given as a product of exponentials of $H_2$ and $H_1$ as in Eq.  \eqref{eq:quench}, it is trivial to compute the transition probabilities, given in Eq.  \eqref{eq:transitionprobabilities}. Since the states appearing as the expressions for $P,P'$, and $P''$ are eigenstates of $H_1$ and $H_2$, and consequently, of $H_{int}$ [see Eq. \eqref{eq:HextHint}], the matrix elements reduce to the scalar product between the instantaneous eigenstates at the beginning and end of the driving strokes, namely,
\begin{eqnarray}
    P &=&\abs{\braket{E_2^{(2)}}{E_2^{(1)}}}^2\,, 
    \nonumber \\
    P' &=&\abs{\braket{E_3^{(2)}}{E_3^{(1)}}}^2\,, \\
    P'' &=&\abs{\braket{E_4^{(2)}}{E_4^{(1)}}}^2
    \nonumber\,,
\end{eqnarray}
where the upper indexes $1,2$ label the initial and final instants of the stroke, and the lower indexes $2,3,4$ label the states with energies $\hbar\omega(t)$, $0$ and $-\hbar\omega(t)$, respectively.

With the expressions for the initial and final Hamiltonian we can easily obtain its eigenstates and compute the inner products above. If we insert these values in the expressions for the friction heats we get
\begin{eqnarray}
    \!\!\!\!\!\!\!\!
    Q_h^{(f)} &=& -(2-\sqrt{2})\,\,\frac{\hbar\omega_1\,\sinh{(\beta_c \hbar \omega_2)}}{1+e^{8 J \beta _c}+2\cosh{(\beta_c\hbar\omega_2)}}\,,
    \nonumber\\
    \!\!\!\!\!\!\!\!
    Q_c^{(f)} &=& -(2-\sqrt{2})\,\,\frac{\hbar\omega_2\,\sinh{(\beta_h\hbar\omega_1)}}{1+e^{8 J \beta _h}+2\cosh{(\beta_h\hbar\omega_1)}}\,.
\end{eqnarray}
The total friction heat is the sum of both contributions $Q^{(f)}=Q_h^{(f)}+Q_c^{(f)}\,$. Notice that the friction heat from both sources is negative, which makes physical sense, since they represent energy losses. It is important to notice that both friction heats have exactly the same functional form if we replace $\omega_1\leftrightarrow\omega_2$ and $T_h\leftrightarrow T_c$.

\begin{figure}[ht]  \centering
\includegraphics[scale=0.35]{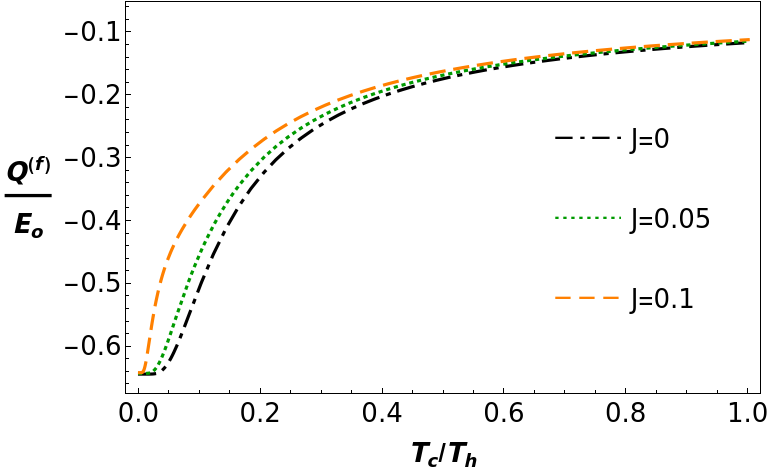}
\includegraphics[scale=0.35]{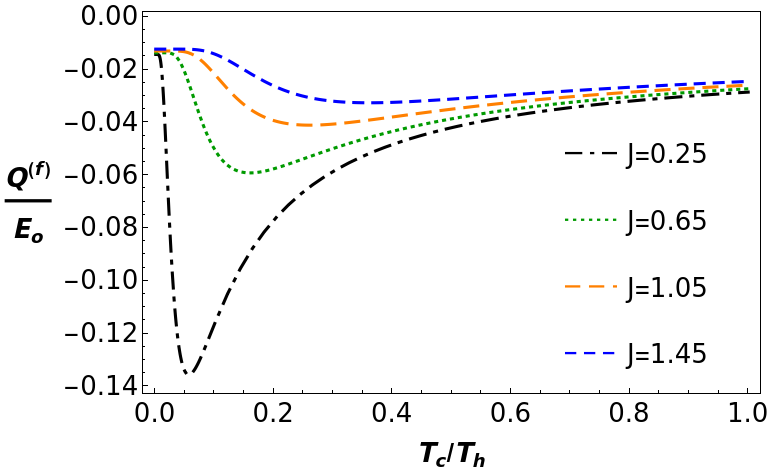} 
\caption{Friction heat as a function of the ratio between temperatures for different values of the coupling. The values used for the Rabi frequencies are $\hbar\omega_1/E_0=2$ and $\hbar\omega_2/E_0=1$. Top panel: In the weak  coupling regime, the friction heat tends to a finite value for small temperatures of the cold source. We have used the value of the temperature of the hot source $k_b T_h/E_0=5\,$. Bottom panel: For the strong coupling regime we have a different behavior. For small values of the temperature of the cold source the engine operates with reduced friction. We have used $k_b T_h/E_0=20\,$.}
\label{fig:fig7}
\end{figure}

By a quick inspection of the expressions above we can see that
the friction heat can behave very differently in the strong and weak coupling regimes, specially in the asymptotic limit of low temperature  ($\beta\rightarrow\infty)$. If $8J>\hbar\omega(t)$, the exponential $e^{8J\beta}$ in the denominator will dominate over the other terms, and the whole quantity will tend to zero at low temperatures. Conversely, if $8J<\hbar\omega(t)$, the friction heat will tend to a finite value for low temperatures. We can see that the engine in the strong coupling regime, although being less efficient in general, can operate with reduced friction in the limit of low temperatures in comparison with the weak coupling regime. In Fig. \ref{fig:fig7}, we plot the friction  heat as a function of the ratio between the bath temperatures $T_c/T_h$ for different values of the coupling in both regimes to illustrate this behavior.

The physical interpretation of this result can be given by the same reasoning as the one presented for the behavior of the efficiency. As we have seen, in the thermalization with the cold source for $T_c\rightarrow 0$, the populations become very concentrated on the ground state, which is the time-independent level  $E_1=-8J$ (singlet state), in the strong coupling regime. Moreover, we have seen in Sec. \ref{sec:TEO} that the dynamics of the working substance does not induce transitions between the singlet and the other eigenstates of the Hamiltonian, due to the isotropy of the interaction between the spins. Since the effects of quantum friction are directly caused by nonadiabatic transitions between the instantaneous energy eigenstates, the supression of these transitions caused by the population of the singlet state will inevitably diminish the overall friction, and that is the effect we are observing here. 

\begin{figure}[h]  \centering
\includegraphics[scale=0.35]{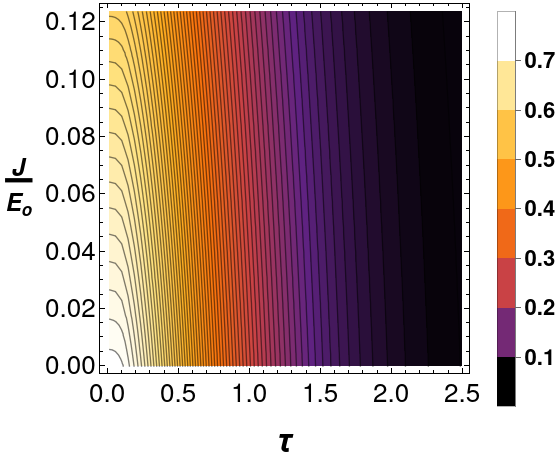}
\includegraphics[scale=0.35]{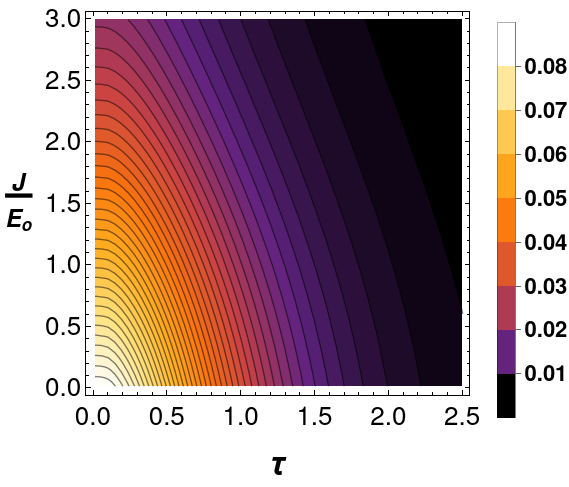} 
\caption{Absolute value of the friction heat $Q^{(f)}$ as a function of the adiabatic parameter and the coupling for the weak (upper) and strong (bottom) regimes. 
The parameters used are $\hbar\omega_1/E_0=2$ and $\hbar\omega_2/E_0=1$ for both regimes, $k_b T_h/E_0=5$ and $T_c/T_h=0.2$ for the weak coupling, and $k_b T_h/E_0=50$ and $T_c/T_h=0.15$ for the strong coupling. In both regimes the friction heat decreases monotonically with both the coupling and the adiabatic parameter.}
\label{fig:fig8}
\end{figure}

It is also interesting to show how the friction heat behaves for protocols that are in the intermediate regime between the quench and adiabatic limits. For this, we return to the hyperbolic tangent protocol described in Sec. \ref{sec:TEO}, and again we use the algebraic methods of Sec. \ref{subsection:NewBCH} for the numerical calculations. In Fig. \ref{fig:fig8} we plot the total friction heat as a function of the adiabatic parameter $\tau$ and the coupling parameter in both weak and strong coupling regimes. We can see that in both regimes the behavior of the friction heat is similar, as they both decrease monotonically with the adiabatic parameter $\tau$. As $\tau$ increases, the evolution becomes more adiabatic and the transitions between the instantaneous energy eigenstates diminish, which leads to a decrease in the losses by quantum friction. 

\section{Conclusion and Outlook}
\label{section:conclusion}

In this work we use an algebraic approach to solve the quantum evolution of the working substance 
in a two-qubit Otto cycle. We investigate the thermodynamic properties of an engine 
formed by the two qubits coupled to each other through an isotropic Heisenberg Hamiltonian 
and with an external magnetic field. Thermal interaction occurs with two heat baths at different 
temperatures, while work is delivered by the external time-dependent magnetic field that varies 
both in amplitude and direction. Different settings of the reservoir temperatures, coupling strength 
and magnetic field amplitude are considered, giving rise to different operation regimes. Our approach 
allows for the investigation of the corresponding operating regions of the engine and its efficiency under 
different protocols determined by the time variation of the external magnetic field.
We show how the coupling and the nonadiabatic driving of the working substance can affect the efficiency,  the output work and the operation modes of the machine. Depending on the cycle parameters, the machine can operate as an 
engine, accelerator, heater, or refrigerator. We have shown how the machine can operate quite differently if the coupling is weak or strong in comparison with the energy associated with the intensity of the driving field, and we have also shown how the machine can operate with reduced friction for high values of the coupling parameter.

In the remainder of this section, we mention other possible analyses that could benefit from the algebraic methods used here.

\subsection{The definition of quantum work and heat}

From a more fundamental consideration, one may explore other definitions of work and heat. In a recent work \cite{Bayesian1} the authors have used Bayesian networks, a well-known concept in statistics and computer science, to define quantum fluctuations relations for the heat exchange that take the initial and final coherence, as well as correlations, into account. Moreover, these results have been experimentally verified \cite{Bayesian2}. This more general definition of heat is more suitable for small interacting systems, and this investigation is a good lead for future projects. 

\subsection{Entropy Considerations}

In the analysis of classical thermal machines, the entropy change in a given cycle is important for dissipative processes, since irreversible protocols produce additional entropy by driving the system beyond the relaxation rate.
The so called information thermodynamics perspective has been used in the study of quantum heat engines \cite{informationengine}. In Ref. \cite{quantumfriction3}, the irreversible work of the system is associated with the aforementioned ``\textit{quantum  friction}", which can be caused by the noncommutativity of the Hamiltonian at different times, and is associated with the production of entropy in the energy basis. They have also shown that the irreversible work caused by friction can be expressed as the relative entropy between the density operator at the end of the work protocol and the density operator at the end of the associated adiabatic process. 
It would be interesting to investigate the associated irreversible work in the two-qubit engine, and compare it with the relative entropy, as well as study other consequences of entropy production in general. 

\subsection{Beyond the isotropic interaction}
It can also be interesting to consider other coupling terms than the isotropic interaction chosen 
in Sec. \ref{sec:interactionhamiltonian}, also used in Refs. \cite{2qbitOttoengine1} and \cite{2qbitOttoengine2}. 
The coupling effects of more complicated interaction models can be more pronounced. One of our future goals is to generalize the formalism used here for these other systems.

Another particularity of our description is the choice of the Otto cycle, which considers complete thermalization strokes followed by pure unitary transformations in the work protocols. If we wait for thermalization, the cycle of the machine will take infinite time. For finite-time cycles, it is necessary to perform partial thermalizations, which requires an analysis of the detailed structure of the heat baths, by the means of the theory of open quantum systems \cite{Kosloff-FourStroke}.

Moreover, we can consider some kind of heat transfer during the work strokes, since no system can be truly isolated from its surroundings. One option to avoid this issue is to perform work strokes during time intervals much smaller than the relaxation time of the system. In this case, the working substance does not thermalize with the baths and we can safely consider the dynamics as purely unitary. Because nuclear spins usually interact weakly with thermal baths, they are a good option for experimental implementations of these protocols \cite{CBPF1,CBPF2}.
\subsection{Final remarks}
Even with all the approximations considered here, we have found interesting results concerning the effects of coupling and nonadiabaticity in the operation of quantum thermal machines. It is important to notice that the formalism presented here can be used to a variety of systems such as, for example, more coupled spins with more complex interactions, or a system of harmonic oscillators operating with some restricted energy levels. It is possible to generalize the idea of the planes associated with the zero value of the heats and work to systems with more energy levels.

The field of quantum thermodynamics is highly connected with the development of better, or more efficient, quantum devices, and therefore experimental implementations are much needed. We hope the methods developed here will serve in future works and contribute to the quest for new quantum technologies. 

\section*{Acknowledgments}

The authors acknowledge Thiago R. de Oliveira for
enlightening discussions. Funding was provided by 
Conselho Nacional de Desenvolvimento Cient\'{\i}fico e Tecnol\'ogico (CNPq),
Coordena\c c\~{a}o de Aperfei\c coamento de Pessoal de N\'\i vel Superior (CAPES), 
Funda\c c\~{a}o Carlos Chagas Filho de Amparo \`{a} Pesquisa do Estado do Rio de Janeiro (FAPERJ), 
and Instituto Nacional de Ci\^encia e Tecnologia de Informa\c c\~ao Qu\^antica 
(INCT-IQ 465469/2014-0).

\bibliography{references.bib}

\begin{thebibliography}{28}%
\makeatletter
\providecommand \@ifxundefined [1]{%
 \@ifx{#1\undefined}
}%
\providecommand \@ifnum [1]{%
 \ifnum #1\expandafter \@firstoftwo
 \else \expandafter \@secondoftwo
 \fi
}%
\providecommand \@ifx [1]{%
 \ifx #1\expandafter \@firstoftwo
 \else \expandafter \@secondoftwo
 \fi
}%
\providecommand \natexlab [1]{#1}%
\providecommand \enquote  [1]{``#1''}%
\providecommand \bibnamefont  [1]{#1}%
\providecommand \bibfnamefont [1]{#1}%
\providecommand \citenamefont [1]{#1}%
\providecommand \href@noop [0]{\@secondoftwo}%
\providecommand \href [0]{\begingroup \@sanitize@url \@href}%
\providecommand \@href[1]{\@@startlink{#1}\@@href}%
\providecommand \@@href[1]{\endgroup#1\@@endlink}%
\providecommand \@sanitize@url [0]{\catcode `\\12\catcode `\$12\catcode
  `\&12\catcode `\#12\catcode `\^12\catcode `\_12\catcode `\%12\relax}%
\providecommand \@@startlink[1]{}%
\providecommand \@@endlink[0]{}%
\providecommand \url  [0]{\begingroup\@sanitize@url \@url }%
\providecommand \@url [1]{\endgroup\@href {#1}{\urlprefix }}%
\providecommand \urlprefix  [0]{URL }%
\providecommand \Eprint [0]{\href }%
\providecommand \doibase [0]{https://doi.org/}%
\providecommand \selectlanguage [0]{\@gobble}%
\providecommand \bibinfo  [0]{\@secondoftwo}%
\providecommand \bibfield  [0]{\@secondoftwo}%
\providecommand \translation [1]{[#1]}%
\providecommand \BibitemOpen [0]{}%
\providecommand \bibitemStop [0]{}%
\providecommand \bibitemNoStop [0]{.\EOS\space}%
\providecommand \EOS [0]{\spacefactor3000\relax}%
\providecommand \BibitemShut  [1]{\csname bibitem#1\endcsname}%
\let\auto@bib@innerbib\@empty
\bibitem [{\citenamefont {Callen}(1985)}]{Callen}%
  \BibitemOpen
  \bibfield  {author} {\bibinfo {author} {\bibfnamefont {H.~B.}\ \bibnamefont
  {Callen}},\ }\bibfield  {title} {\bibinfo {title} {\textit{Thermodynamics and
  an introduction to thermostatistics; 2nd ed.}},\ }\href
  {https://cds.cern.ch/record/450289} {\bibfield  {journal} {\bibinfo
  {journal} {Wiley, New York, NY}\ } (\bibinfo {year} {1985})}\BibitemShut
  {NoStop}%
\bibitem [{\citenamefont {Scovil}\ and\ \citenamefont
  {Schulz-DuBois}(1959)}]{3levelmaser}%
  \BibitemOpen
  \bibfield  {author} {\bibinfo {author} {\bibfnamefont {H.~E.~D.}\
  \bibnamefont {Scovil}}\ and\ \bibinfo {author} {\bibfnamefont {E.~O.}\
  \bibnamefont {Schulz-DuBois}},\ }\bibfield  {title} {\bibinfo {title}
  {Three-level masers as heat engines},\ }\href
  {https://doi.org/10.1103/PhysRevLett.2.262} {\bibfield  {journal} {\bibinfo
  {journal} {Phys. Rev. Lett.}\ }\textbf {\bibinfo {volume} {2}},\ \bibinfo
  {pages} {262} (\bibinfo {year} {1959})}\BibitemShut {NoStop}%
\bibitem [{\citenamefont {Batalh\~ao}\ \emph {et~al.}(2014)\citenamefont
  {Batalh\~ao}, \citenamefont {Souza}, \citenamefont {Mazzola}, \citenamefont
  {Auccaise}, \citenamefont {Sarthour}, \citenamefont {Oliveira}, \citenamefont
  {Goold}, \citenamefont {De~Chiara}, \citenamefont {Paternostro},\ and\
  \citenamefont {Serra}}]{CBPF1}%
  \BibitemOpen
  \bibfield  {author} {\bibinfo {author} {\bibfnamefont {T.~B.}\ \bibnamefont
  {Batalh\~ao}}, \bibinfo {author} {\bibfnamefont {A.~M.}\ \bibnamefont
  {Souza}}, \bibinfo {author} {\bibfnamefont {L.}~\bibnamefont {Mazzola}},
  \bibinfo {author} {\bibfnamefont {R.}~\bibnamefont {Auccaise}}, \bibinfo
  {author} {\bibfnamefont {R.~S.}\ \bibnamefont {Sarthour}}, \bibinfo {author}
  {\bibfnamefont {I.~S.}\ \bibnamefont {Oliveira}}, \bibinfo {author}
  {\bibfnamefont {J.}~\bibnamefont {Goold}}, \bibinfo {author} {\bibfnamefont
  {G.}~\bibnamefont {De~Chiara}}, \bibinfo {author} {\bibfnamefont
  {M.}~\bibnamefont {Paternostro}},\ and\ \bibinfo {author} {\bibfnamefont
  {R.~M.}\ \bibnamefont {Serra}},\ }\bibfield  {title} {\bibinfo {title}
  {Experimental reconstruction of work distribution and study of fluctuation
  relations in a closed quantum system},\ }\href
  {https://doi.org/10.1103/PhysRevLett.113.140601} {\bibfield  {journal}
  {\bibinfo  {journal} {Phys. Rev. Lett.}\ }\textbf {\bibinfo {volume} {113}},\
  \bibinfo {pages} {140601} (\bibinfo {year} {2014})}\BibitemShut {NoStop}%
\bibitem [{\citenamefont {Peterson}\ \emph {et~al.}(2019)\citenamefont
  {Peterson}, \citenamefont {Batalh\~ao}, \citenamefont {Herrera},
  \citenamefont {Souza}, \citenamefont {Sarthour}, \citenamefont {Oliveira},\
  and\ \citenamefont {Serra}}]{CBPF2}%
  \BibitemOpen
  \bibfield  {author} {\bibinfo {author} {\bibfnamefont {J.~P.~S.}\
  \bibnamefont {Peterson}}, \bibinfo {author} {\bibfnamefont {T.~B.}\
  \bibnamefont {Batalh\~ao}}, \bibinfo {author} {\bibfnamefont
  {M.}~\bibnamefont {Herrera}}, \bibinfo {author} {\bibfnamefont {A.~M.}\
  \bibnamefont {Souza}}, \bibinfo {author} {\bibfnamefont {R.~S.}\ \bibnamefont
  {Sarthour}}, \bibinfo {author} {\bibfnamefont {I.~S.}\ \bibnamefont
  {Oliveira}},\ and\ \bibinfo {author} {\bibfnamefont {R.~M.}\ \bibnamefont
  {Serra}},\ }\bibfield  {title} {\bibinfo {title} {Experimental
  characterization of a spin quantum heat engine},\ }\href
  {https://doi.org/10.1103/PhysRevLett.123.240601} {\bibfield  {journal}
  {\bibinfo  {journal} {Phys. Rev. Lett.}\ }\textbf {\bibinfo {volume} {123}},\
  \bibinfo {pages} {240601} (\bibinfo {year} {2019})}\BibitemShut {NoStop}%
\bibitem [{\citenamefont {Feldmann}\ and\ \citenamefont
  {Kosloff}(2003)}]{Kosloff-FourStroke}%
  \BibitemOpen
  \bibfield  {author} {\bibinfo {author} {\bibfnamefont {T.}~\bibnamefont
  {Feldmann}}\ and\ \bibinfo {author} {\bibfnamefont {R.}~\bibnamefont
  {Kosloff}},\ }\bibfield  {title} {\bibinfo {title} {Quantum four-stroke heat
  engine: Thermodynamic observables in a model with intrinsic friction},\
  }\href {https://doi.org/10.1103/PhysRevE.68.016101} {\bibfield  {journal}
  {\bibinfo  {journal} {Phys. Rev. E}\ }\textbf {\bibinfo {volume} {68}},\
  \bibinfo {pages} {016101} (\bibinfo {year} {2003})}\BibitemShut {NoStop}%
\bibitem [{\citenamefont {Türkpençe}\ and\ \citenamefont
  {Altintas}(2019{\natexlab{a}})}]{article}%
  \BibitemOpen
  \bibfield  {author} {\bibinfo {author} {\bibfnamefont {D.}~\bibnamefont
  {Türkpençe}}\ and\ \bibinfo {author} {\bibfnamefont {F.}~\bibnamefont
  {Altintas}},\ }\bibfield  {title} {\bibinfo {title} {Coupled quantum otto
  heat engine and refrigerator with inner friction},\ }\href
  {https://doi.org/10.1007/s11128-019-2366-7} {\bibfield  {journal} {\bibinfo
  {journal} {Quantum Information Processing}\ }\textbf {\bibinfo {volume} {18}}
  (\bibinfo {year} {2019}{\natexlab{a}})}\BibitemShut {NoStop}%
\bibitem [{\citenamefont {Martínez-Tibaduiza}\ \emph
  {et~al.}(2020)\citenamefont {Martínez-Tibaduiza}, \citenamefont {Aragão},
  \citenamefont {Farina},\ and\ \citenamefont {Zarro}}]{NewBCH}%
  \BibitemOpen
  \bibfield  {author} {\bibinfo {author} {\bibfnamefont {D.}~\bibnamefont
  {Martínez-Tibaduiza}}, \bibinfo {author} {\bibfnamefont {A.}~\bibnamefont
  {Aragão}}, \bibinfo {author} {\bibfnamefont {C.}~\bibnamefont {Farina}},\
  and\ \bibinfo {author} {\bibfnamefont {C.}~\bibnamefont {Zarro}},\ }\bibfield
   {title} {\bibinfo {title} {New {BCH}-like relations of the su(1,1), su(2)
  and so(2,1) {L}ie algebras},\ }\href
  {https://doi.org/https://doi.org/10.1016/j.physleta.2020.126937} {\bibfield
  {journal} {\bibinfo  {journal} {Physics Letters A}\ }\textbf {\bibinfo
  {volume} {384}},\ \bibinfo {pages} {126937} (\bibinfo {year}
  {2020})}\BibitemShut {NoStop}%
\bibitem [{\citenamefont {Solfanelli}\ \emph {et~al.}(2020)\citenamefont
  {Solfanelli}, \citenamefont {Falsetti},\ and\ \citenamefont
  {Campisi}}]{Campisi1}%
  \BibitemOpen
  \bibfield  {author} {\bibinfo {author} {\bibfnamefont {A.}~\bibnamefont
  {Solfanelli}}, \bibinfo {author} {\bibfnamefont {M.}~\bibnamefont
  {Falsetti}},\ and\ \bibinfo {author} {\bibfnamefont {M.}~\bibnamefont
  {Campisi}},\ }\bibfield  {title} {\bibinfo {title} {Nonadiabatic single-qubit
  quantum otto engine},\ }\bibfield  {journal} {\bibinfo  {journal} {Physical
  Review B}\ }\textbf {\bibinfo {volume} {101}},\ \href
  {https://doi.org/10.1103/physrevb.101.054513} {10.1103/physrevb.101.054513}
  (\bibinfo {year} {2020})\BibitemShut {NoStop}%
\bibitem [{\citenamefont {Thomas}\ and\ \citenamefont
  {Johal}(2011)}]{2qbitOttoengine1}%
  \BibitemOpen
  \bibfield  {author} {\bibinfo {author} {\bibfnamefont {G.}~\bibnamefont
  {Thomas}}\ and\ \bibinfo {author} {\bibfnamefont {R.~S.}\ \bibnamefont
  {Johal}},\ }\bibfield  {title} {\bibinfo {title} {Coupled quantum otto
  cycle},\ }\href {https://doi.org/10.1103/PhysRevE.83.031135} {\bibfield
  {journal} {\bibinfo  {journal} {Phys. Rev. E}\ }\textbf {\bibinfo {volume}
  {83}},\ \bibinfo {pages} {031135} (\bibinfo {year} {2011})}\BibitemShut
  {NoStop}%
\bibitem [{\citenamefont {Radmore}\ and\ \citenamefont
  {Barnett}(1997)}]{Barnett-Book-1997}%
  \BibitemOpen
  \bibfield  {author} {\bibinfo {author} {\bibfnamefont {P.~M.}\ \bibnamefont
  {Radmore}}\ and\ \bibinfo {author} {\bibfnamefont {S.~M.}\ \bibnamefont
  {Barnett}},\ }\bibfield  {title} {\bibinfo {title} {\textit{Methods in
  theoretical quantum optics}},\ }\href@noop {} {\bibfield  {journal} {\bibinfo
   {journal} {Cambridge University Press}\ } (\bibinfo {year}
  {1997})}\BibitemShut {NoStop}%
\bibitem [{\citenamefont {Gilmore}(2005)}]{gilmore2012lie}%
  \BibitemOpen
  \bibfield  {author} {\bibinfo {author} {\bibfnamefont {R.}~\bibnamefont
  {Gilmore}},\ }\href@noop {} {\emph {\bibinfo {title} {\textit{Lie groups, Lie
  algebras, and some of their applications}}}}\ (\bibinfo  {publisher} {Dover
  Publications},\ \bibinfo {address} {Mineola, New York},\ \bibinfo {year}
  {2005})\BibitemShut {NoStop}%
\bibitem [{\citenamefont {Martínez~Tibaduiza}\ \emph
  {et~al.}(2020)\citenamefont {Martínez~Tibaduiza}, \citenamefont {Pires},
  \citenamefont {Szilard}, \citenamefont {Zarro}, \citenamefont {Farina~de
  Souza},\ and\ \citenamefont {Rego}}]{twojumps}%
  \BibitemOpen
  \bibfield  {author} {\bibinfo {author} {\bibfnamefont {D.}~\bibnamefont
  {Martínez~Tibaduiza}}, \bibinfo {author} {\bibfnamefont {L.}~\bibnamefont
  {Pires}}, \bibinfo {author} {\bibfnamefont {D.}~\bibnamefont {Szilard}},
  \bibinfo {author} {\bibfnamefont {C.}~\bibnamefont {Zarro}}, \bibinfo
  {author} {\bibfnamefont {C.}~\bibnamefont {Farina~de Souza}},\ and\ \bibinfo
  {author} {\bibfnamefont {A.}~\bibnamefont {Rego}},\ }\bibfield  {title}
  {\bibinfo {title} {A time-dependent harmonic oscillator with two frequency
  jumps: an exact algebraic solution},\ }\href
  {https://doi.org/10.1007/s13538-020-00770-x} {\bibfield  {journal} {\bibinfo
  {journal} {Brazilian Journal of Physics}\ }\textbf {\bibinfo {volume} {50}}
  (\bibinfo {year} {2020})}\BibitemShut {NoStop}%
\bibitem [{\citenamefont {Khinchin}\ and\ \citenamefont
  {Eagle}(1997)}]{khinchin1997continued}%
  \BibitemOpen
  \bibfield  {author} {\bibinfo {author} {\bibfnamefont {A.}~\bibnamefont
  {Khinchin}}\ and\ \bibinfo {author} {\bibfnamefont {H.}~\bibnamefont
  {Eagle}},\ }\bibfield  {title} {\bibinfo {title} {\textit{Continued
  Fractions}},\ }\href {https://books.google.com.br/books?id=R7Fp8vytgeAC}
  {\bibfield  {journal} {\bibinfo  {journal} {Dover books on mathematics}\ }
  (\bibinfo {year} {1997})}\BibitemShut {NoStop}%
\bibitem [{\citenamefont {Wei}\ and\ \citenamefont {Norman}(1963)}]{Wei_1963}%
  \BibitemOpen
  \bibfield  {author} {\bibinfo {author} {\bibfnamefont {J.}~\bibnamefont
  {Wei}}\ and\ \bibinfo {author} {\bibfnamefont {E.}~\bibnamefont {Norman}},\
  }\bibfield  {title} {\bibinfo {title} {Lie algebraic solution of linear
  differential equations},\ }\href {https://doi.org/10.1063/1.1703993}
  {\bibfield  {journal} {\bibinfo  {journal} {Journal of Mathematical Physics}\
  }\textbf {\bibinfo {volume} {4}},\ \bibinfo {pages} {575} (\bibinfo {year}
  {1963})}\BibitemShut {NoStop}%
\bibitem [{\citenamefont {Wei}\ and\ \citenamefont {Norman}(1964)}]{Wei_1964}%
  \BibitemOpen
  \bibfield  {author} {\bibinfo {author} {\bibfnamefont {J.}~\bibnamefont
  {Wei}}\ and\ \bibinfo {author} {\bibfnamefont {E.}~\bibnamefont {Norman}},\
  }\bibfield  {title} {\bibinfo {title} {On global representations of the
  solutions of linear differential equations as a product of exponentials},\
  }\href {https://doi.org/https://doi.org/10.2307/2034065} {\bibfield
  {journal} {\bibinfo  {journal} {Proceedings of the American Mathematical
  Society}\ }\textbf {\bibinfo {volume} {15}},\ \bibinfo {pages} {327}
  (\bibinfo {year} {1964})}\BibitemShut {NoStop}%
\bibitem [{\citenamefont {Tibaduiza}\ \emph {et~al.}(2020)\citenamefont
  {Tibaduiza}, \citenamefont {Pires}, \citenamefont {Rego}, \citenamefont
  {Szilard}, \citenamefont {Zarro},\ and\ \citenamefont
  {Farina}}]{efficientsolutionTDHO}%
  \BibitemOpen
  \bibfield  {author} {\bibinfo {author} {\bibfnamefont {D.~M.}\ \bibnamefont
  {Tibaduiza}}, \bibinfo {author} {\bibfnamefont {L.}~\bibnamefont {Pires}},
  \bibinfo {author} {\bibfnamefont {A.~L.~C.}\ \bibnamefont {Rego}}, \bibinfo
  {author} {\bibfnamefont {D.}~\bibnamefont {Szilard}}, \bibinfo {author}
  {\bibfnamefont {C.}~\bibnamefont {Zarro}},\ and\ \bibinfo {author}
  {\bibfnamefont {C.}~\bibnamefont {Farina}},\ }\bibfield  {title} {\bibinfo
  {title} {Efficient algebraic solution for a time-dependent quantum harmonic
  oscillator},\ }\href {https://doi.org/10.1088/1402-4896/abb254} {\bibfield
  {journal} {\bibinfo  {journal} {Physica Scripta}\ }\textbf {\bibinfo {volume}
  {95}},\ \bibinfo {pages} {105102} (\bibinfo {year} {2020})}\BibitemShut
  {NoStop}%
\bibitem [{\citenamefont {Mart{\'\i}nez-Tibaduiza}\ \emph
  {et~al.}(2021)\citenamefont {Mart{\'\i}nez-Tibaduiza}, \citenamefont
  {Pires},\ and\ \citenamefont {Farina}}]{DMT-JPHYSB-2021}%
  \BibitemOpen
  \bibfield  {author} {\bibinfo {author} {\bibfnamefont {D.}~\bibnamefont
  {Mart{\'\i}nez-Tibaduiza}}, \bibinfo {author} {\bibfnamefont
  {L.}~\bibnamefont {Pires}},\ and\ \bibinfo {author} {\bibfnamefont
  {C.}~\bibnamefont {Farina}},\ }\bibfield  {title} {\bibinfo {title}
  {Time-dependent quantum harmonic oscillator: a continuous route from
  adiabatic to sudden changes},\ }\href@noop {} {\bibfield  {journal} {\bibinfo
   {journal} {Journal of Physics B: Atomic, Molecular and Optical Physics}\
  }\textbf {\bibinfo {volume} {54}},\ \bibinfo {pages} {205401} (\bibinfo
  {year} {2021})}\BibitemShut {NoStop}%
\bibitem [{\citenamefont {Šamaj}\ and\ \citenamefont
  {Bajnok}(2013)}]{Statistical_Many_Body}%
  \BibitemOpen
  \bibfield  {author} {\bibinfo {author} {\bibfnamefont {L.}~\bibnamefont
  {Šamaj}}\ and\ \bibinfo {author} {\bibfnamefont {Z.}~\bibnamefont
  {Bajnok}},\ }\bibfield  {title} {\bibinfo {title} {\textit{Introduction to
  the Statistical Physics of Integrable Many-body Systems}},\ }\bibfield
  {journal} {\bibinfo  {journal} {Cambridge University Press}\ }\href
  {https://doi.org/10.1017/CBO9781139343480} {10.1017/CBO9781139343480}
  (\bibinfo {year} {2013})\BibitemShut {NoStop}%
\bibitem [{\citenamefont {de~Oliveira}\ and\ \citenamefont
  {Jonathan}(2021)}]{OttoCycleUFF}%
  \BibitemOpen
  \bibfield  {author} {\bibinfo {author} {\bibfnamefont {T.~R.}\ \bibnamefont
  {de~Oliveira}}\ and\ \bibinfo {author} {\bibfnamefont {D.}~\bibnamefont
  {Jonathan}},\ }\bibfield  {title} {\bibinfo {title} {Efficiency gain and
  bidirectional operation of quantum engines with decoupled internal levels},\
  }\href {https://doi.org/10.1103/PhysRevE.104.044133} {\bibfield  {journal}
  {\bibinfo  {journal} {Phys. Rev. E}\ }\textbf {\bibinfo {volume} {104}},\
  \bibinfo {pages} {044133} (\bibinfo {year} {2021})}\BibitemShut {NoStop}%
\bibitem [{\citenamefont {Cohen-Tannoudji}\ \emph {et~al.}(1977)\citenamefont
  {Cohen-Tannoudji}, \citenamefont {Diu},\ and\ \citenamefont
  {Laloë}}]{Cohen-Tannoudji}%
  \BibitemOpen
  \bibfield  {author} {\bibinfo {author} {\bibfnamefont {C.}~\bibnamefont
  {Cohen-Tannoudji}}, \bibinfo {author} {\bibfnamefont {B.}~\bibnamefont
  {Diu}},\ and\ \bibinfo {author} {\bibfnamefont {F.}~\bibnamefont {Laloë}},\
  }\bibfield  {title} {\bibinfo {title} {{\textit{Quantum mechanics; 1st
  ed.}}},\ }\href {https://cds.cern.ch/record/101367} {\bibfield  {journal}
  {\bibinfo  {journal} {Wiley, New York, NY}\ } (\bibinfo {year} {1977})},\
  \bibinfo {note} {trans. of : Mécanique quantique. Paris : Hermann,
  1973}\BibitemShut {NoStop}%
\bibitem [{\citenamefont {Zhang}\ \emph {et~al.}(2007)\citenamefont {Zhang},
  \citenamefont {Liu}, \citenamefont {Chen},\ and\ \citenamefont
  {Li}}]{2qbitOttoengine2}%
  \BibitemOpen
  \bibfield  {author} {\bibinfo {author} {\bibfnamefont {T.}~\bibnamefont
  {Zhang}}, \bibinfo {author} {\bibfnamefont {W.-T.}\ \bibnamefont {Liu}},
  \bibinfo {author} {\bibfnamefont {P.-X.}\ \bibnamefont {Chen}},\ and\
  \bibinfo {author} {\bibfnamefont {C.-Z.}\ \bibnamefont {Li}},\ }\bibfield
  {title} {\bibinfo {title} {Four-level entangled quantum heat engines},\
  }\href {https://doi.org/10.1103/PhysRevA.75.062102} {\bibfield  {journal}
  {\bibinfo  {journal} {Phys. Rev. A}\ }\textbf {\bibinfo {volume} {75}},\
  \bibinfo {pages} {062102} (\bibinfo {year} {2007})}\BibitemShut {NoStop}%
\bibitem [{\citenamefont {Altintas}\ and\ \citenamefont {Özgür
  E.~Müstecaplıoğlu}(2015)}]{2qbitOttoengine3}%
  \BibitemOpen
  \bibfield  {author} {\bibinfo {author} {\bibfnamefont {F.}~\bibnamefont
  {Altintas}}\ and\ \bibinfo {author} {\bibnamefont {Özgür
  E.~Müstecaplıoğlu}},\ }\bibfield  {title} {\bibinfo {title} {General
  formalism of local thermodynamics with an example: Quantum otto engine with a
  spin-$1/2$ coupled to an arbitrary spin},\ }\href
  {https://doi.org/10.1103/PhysRevE.92.022142} {\bibfield  {journal} {\bibinfo
  {journal} {Phys. Rev. E}\ }\textbf {\bibinfo {volume} {92}},\ \bibinfo
  {pages} {022142} (\bibinfo {year} {2015})}\BibitemShut {NoStop}%
\bibitem [{\citenamefont {Türkpençe}\ and\ \citenamefont
  {Altintas}(2019{\natexlab{b}})}]{quantumfriction1}%
  \BibitemOpen
  \bibfield  {author} {\bibinfo {author} {\bibfnamefont {D.}~\bibnamefont
  {Türkpençe}}\ and\ \bibinfo {author} {\bibfnamefont {F.}~\bibnamefont
  {Altintas}},\ }\bibfield  {title} {\bibinfo {title} {Coupled quantum otto
  heat engine and refrigerator with inner friction},\ }\href
  {https://doi.org/10.1007/s11128-019-2366-7} {\bibfield  {journal} {\bibinfo
  {journal} {Quantum Information Processing}\ }\textbf {\bibinfo {volume} {18}}
  (\bibinfo {year} {2019}{\natexlab{b}})}\BibitemShut {NoStop}%
\bibitem [{\citenamefont {Plastina}\ \emph
  {et~al.}(2014{\natexlab{a}})\citenamefont {Plastina}, \citenamefont {Alecce},
  \citenamefont {Apollaro}, \citenamefont {Falcone}, \citenamefont {Francica},
  \citenamefont {Galve}, \citenamefont {Lo~Gullo},\ and\ \citenamefont
  {Zambrini}}]{quantumfriction2}%
  \BibitemOpen
  \bibfield  {author} {\bibinfo {author} {\bibfnamefont {F.}~\bibnamefont
  {Plastina}}, \bibinfo {author} {\bibfnamefont {A.}~\bibnamefont {Alecce}},
  \bibinfo {author} {\bibfnamefont {T.~J.~G.}\ \bibnamefont {Apollaro}},
  \bibinfo {author} {\bibfnamefont {G.}~\bibnamefont {Falcone}}, \bibinfo
  {author} {\bibfnamefont {G.}~\bibnamefont {Francica}}, \bibinfo {author}
  {\bibfnamefont {F.}~\bibnamefont {Galve}}, \bibinfo {author} {\bibfnamefont
  {N.}~\bibnamefont {Lo~Gullo}},\ and\ \bibinfo {author} {\bibfnamefont
  {R.}~\bibnamefont {Zambrini}},\ }\bibfield  {title} {\bibinfo {title}
  {Irreversible work and inner friction in quantum thermodynamic processes},\
  }\href {https://doi.org/10.1103/PhysRevLett.113.260601} {\bibfield  {journal}
  {\bibinfo  {journal} {Phys. Rev. Lett.}\ }\textbf {\bibinfo {volume} {113}},\
  \bibinfo {pages} {260601} (\bibinfo {year} {2014}{\natexlab{a}})}\BibitemShut
  {NoStop}%
\bibitem [{\citenamefont {Micadei}\ \emph {et~al.}(2020)\citenamefont
  {Micadei}, \citenamefont {Landi},\ and\ \citenamefont {Lutz}}]{Bayesian1}%
  \BibitemOpen
  \bibfield  {author} {\bibinfo {author} {\bibfnamefont {K.}~\bibnamefont
  {Micadei}}, \bibinfo {author} {\bibfnamefont {G.~T.}\ \bibnamefont {Landi}},\
  and\ \bibinfo {author} {\bibfnamefont {E.}~\bibnamefont {Lutz}},\ }\bibfield
  {title} {\bibinfo {title} {Quantum fluctuation theorems beyond two-point
  measurements},\ }\href {https://doi.org/10.1103/PhysRevLett.124.090602}
  {\bibfield  {journal} {\bibinfo  {journal} {Phys. Rev. Lett.}\ }\textbf
  {\bibinfo {volume} {124}},\ \bibinfo {pages} {090602} (\bibinfo {year}
  {2020})}\BibitemShut {NoStop}%
\bibitem [{\citenamefont {Micadei}\ \emph {et~al.}(2021)\citenamefont
  {Micadei}, \citenamefont {Peterson}, \citenamefont {Souza}, \citenamefont
  {Sarthour}, \citenamefont {Oliveira}, \citenamefont {Landi}, \citenamefont
  {Serra},\ and\ \citenamefont {Lutz}}]{Bayesian2}%
  \BibitemOpen
  \bibfield  {author} {\bibinfo {author} {\bibfnamefont {K.}~\bibnamefont
  {Micadei}}, \bibinfo {author} {\bibfnamefont {J.~P.~S.}\ \bibnamefont
  {Peterson}}, \bibinfo {author} {\bibfnamefont {A.~M.}\ \bibnamefont {Souza}},
  \bibinfo {author} {\bibfnamefont {R.~S.}\ \bibnamefont {Sarthour}}, \bibinfo
  {author} {\bibfnamefont {I.~S.}\ \bibnamefont {Oliveira}}, \bibinfo {author}
  {\bibfnamefont {G.~T.}\ \bibnamefont {Landi}}, \bibinfo {author}
  {\bibfnamefont {R.~M.}\ \bibnamefont {Serra}},\ and\ \bibinfo {author}
  {\bibfnamefont {E.}~\bibnamefont {Lutz}},\ }\bibfield  {title} {\bibinfo
  {title} {Experimental validation of fully quantum fluctuation theorems using
  dynamic bayesian networks},\ }\href
  {https://doi.org/10.1103/PhysRevLett.127.180603} {\bibfield  {journal}
  {\bibinfo  {journal} {Phys. Rev. Lett.}\ }\textbf {\bibinfo {volume} {127}},\
  \bibinfo {pages} {180603} (\bibinfo {year} {2021})}\BibitemShut {NoStop}%
\bibitem [{\citenamefont {Zhou}\ and\ \citenamefont
  {Segal}(2010)}]{informationengine}%
  \BibitemOpen
  \bibfield  {author} {\bibinfo {author} {\bibfnamefont {Y.}~\bibnamefont
  {Zhou}}\ and\ \bibinfo {author} {\bibfnamefont {D.}~\bibnamefont {Segal}},\
  }\bibfield  {title} {\bibinfo {title} {Minimal model of a heat engine:
  Information theory approach},\ }\href
  {https://doi.org/10.1103/PhysRevE.82.011120} {\bibfield  {journal} {\bibinfo
  {journal} {Phys. Rev. E}\ }\textbf {\bibinfo {volume} {82}},\ \bibinfo
  {pages} {011120} (\bibinfo {year} {2010})}\BibitemShut {NoStop}%
\bibitem [{\citenamefont {Plastina}\ \emph
  {et~al.}(2014{\natexlab{b}})\citenamefont {Plastina}, \citenamefont {Alecce},
  \citenamefont {Apollaro}, \citenamefont {Falcone}, \citenamefont {Francica},
  \citenamefont {Galve}, \citenamefont {Lo~Gullo},\ and\ \citenamefont
  {Zambrini}}]{quantumfriction3}%
  \BibitemOpen
  \bibfield  {author} {\bibinfo {author} {\bibfnamefont {F.}~\bibnamefont
  {Plastina}}, \bibinfo {author} {\bibfnamefont {A.}~\bibnamefont {Alecce}},
  \bibinfo {author} {\bibfnamefont {T.~J.~G.}\ \bibnamefont {Apollaro}},
  \bibinfo {author} {\bibfnamefont {G.}~\bibnamefont {Falcone}}, \bibinfo
  {author} {\bibfnamefont {G.}~\bibnamefont {Francica}}, \bibinfo {author}
  {\bibfnamefont {F.}~\bibnamefont {Galve}}, \bibinfo {author} {\bibfnamefont
  {N.}~\bibnamefont {Lo~Gullo}},\ and\ \bibinfo {author} {\bibfnamefont
  {R.}~\bibnamefont {Zambrini}},\ }\bibfield  {title} {\bibinfo {title}
  {Irreversible work and inner friction in quantum thermodynamic processes},\
  }\href {https://doi.org/10.1103/PhysRevLett.113.260601} {\bibfield  {journal}
  {\bibinfo  {journal} {Phys. Rev. Lett.}\ }\textbf {\bibinfo {volume} {113}},\
  \bibinfo {pages} {260601} (\bibinfo {year} {2014}{\natexlab{b}})}\BibitemShut
  {NoStop}%
\end{thebibliography}%

\end{document}